\pdfoutput=1 

\documentclass[10pt,aps,prl,twocolumn,showpacs,nofootinbib,longbibliography]{revtex4-1}

\usepackage{amsmath,amsthm,amssymb,amsfonts,url}
\usepackage{bbm}
\usepackage[usenames,dvipsnames]{color}
\usepackage{hyperref}

\hypersetup{
pdfkeywords={spinfoam, spin foam, loop quantum gravity, Einstein equations, general relativity, Barbero-Immirzi parameter, Regge calculus, classical limit, coherent states}, 
colorlinks=true,         
linkcolor=OliveGreen,          
citecolor=OliveGreen,        
urlcolor=OliveGreen,      
}

\makeatletter
\renewcommand*{\@fnsymbol}[1]{\ensuremath{%
  \ifcase#1\or
   *\or
   **\or
   ***\or
   \mathsection\or
   \mathparagraph\or
   \dagger\or
   \ddagger\or
   \mathsection\mathsection\or
   \mathparagraph\mathparagraph\or
   \mathsection\mathsection\mathsection\or
   \mathparagraph\mathparagraph\mathparagraph\or
   \dagger\dagger\or
   \ddagger\ddagger\else
   \dagger\dagger\dagger\or
   \ddagger\ddagger\ddagger\else
   \@ctrerr
  \fi
 }%
}%
\makeatother

\begin{document}                

\newcommand{\mc}[1]{\mathcal{#1}}
\newcommand{\mbb}[1]{\mathbbm{#1}}
\newcommand{\Tr}{{\rm Tr}}
\newcommand{\ket}[1]{|#1\rangle}
\newcommand{\bra}[1]{\langle#1|}
\newcommand{\braket}[2]{\langle#1|#2\rangle}
\newcommand{\im}[1]{\text{Im}\,#1}
\newcommand{\re}[1]{\text{Re}\,#1}
\newcommand{\comment}[1]{}

\title{\large\bf Regge gravity from spinfoams}

\author{Elena Magliaro}
\email{magliaro@gravity.psu.edu}
\author{Claudio Perini}
\email{perini@gravity.psu.edu}
\affiliation{Institute for Gravitation and the Cosmos, Physics Department, Penn State, University Park, PA 16802-6300, USA}

\date{\small\today}

\begin{abstract}
We consider spinfoam quantum gravity for general triangulations in the regime $l_P^2\ll a\ll a/\gamma$, namely in the combined classical limit of large areas $a$ and flipped limit of small Barbero-Immirzi parameter $\gamma$, where $l_P$ is the Planck length. Under few working hypotheses we find that the flipped limit enforces the constraints that turn the spinfoam theory into an effective Regge-like quantum theory with lengths as variables, while the classical limit selects among the possible geometries the ones satisfying the Einstein equations. Two kinds of quantum corrections appear in terms of powers of $l^2_P/a$ and $\gamma l_P^2/a$. The result also suggests that the Barbero-Immirzi parameter may run to smaller values under coarse-graining of the triangulation. 
\end{abstract}

\maketitle

\section{Introduction} 
Spinfoams \cite{Baez:1997zt,Reisenberger:2000fy,Perez:2003vx} are a tentative covariant quantization of general relativity. They provide transition amplitudes between quantum states of 3-geometry, in the form of a Misner-Hawking sum over virtual 4-geometries \cite{Misner:1957wq,Hawking:1979zw}. They are usually defined on a fixed spacetime triangulation using a discretized version of the classical action of general relativity, and the full degrees of freedom are recovered in the refinement limit of the triangulation \cite{Rovelli:2010qx}.

In the so-called `new models' \cite{Engle:2007wy,Livine:2007ya,Freidel:2007py,Rovelli:2010bf} considered in this paper the intermediate quantum states are the ones of Hamiltonian loop quantum gravity, the $SU(2)$ spin-network states, a remarkable feature that promotes the spinfoam framework to a tentative path integral representation of loop quantum gravity. Moreover, the spinfoam amplitude has been shown to possess the correct local Lorentz symmetries \cite{Rovelli:2010ed} and has been coupled to matter and Yang-Mills fields \cite{Bianchi:2010bn,Han:2011as}.

The Holst formulation of general relativity underlies the construction of loop quantum gravity as well as its covariant version given by spinfoam models. In the vacuum the Holst action reads
\begin{align}\label{Sintro}
S(e,\omega)=\int (e\wedge e)^*\wedge F(\omega)+\frac{1}{\gamma}\int (e\wedge e)\wedge F(\omega),
\end{align}
where the first term is the Einstein-Hilbert action written in the first-order Palatini formalism in terms of a cotetrad $e$ and a  connection $\omega$, and the second term is called the Holst term.\footnote{The cotetrad $e^I_{\mu}dx^\mu$ is a set of four 1-forms related to the metric via $g_{\mu\nu}=e^I_{\mu}e^J_{\nu}\delta_{IJ}$. The cotetrad, or better its inverse $e^\mu_I\partial_\mu$ (the tetrad), defines an inertial frame at each spacetime point. The connection $\omega^{IJ}_\mu dx^\mu$ is a 1-form with values in the Lorentz algebra $so(1,3)$, thus it is antisymmetric in the internal indices. The curvature of the connection is $F(\omega)^{IJ}:=d\omega^{IJ}+\omega^I_{\;K}\wedge\omega^{KJ}$. Notice that in the action \eqref{Sintro} a full contraction of internal indices is understood.} The star $*$ denotes the Hodge duality on the internal indices.

The real parameter $\gamma$ is the Barbero-Immirzi parameter, which does not play any special role in the classical theory. In fact, varying the action with respect to $\omega$ one gets the Cartan equation
\begin{align}\label{Cartan}
T^I:=de^I+\omega^I_{\;J}\wedge e^J=0,
\end{align}
namely the vanishing of the torsion $T$. Thus the Cartan equation forces $\omega$ to be the unique torsion-free connection $\omega=\omega(e)$ (akin to the Levi-Civita connection of the metric formalism) and makes the Holst term vanish. The Einstein equations
\begin{align}\label{Einsteintetrad}
\epsilon_{IJKL} e^J\wedge F(\omega(e))^{KL}=0
\end{align}
are obtained by varying the cotetrad $e$. However, the Barbero-Immirzi parameter is known to affect the quantum theory as it enters the discrete spectra of area and volume operators \cite{Rovelli:1994ge, Ding:2010ye}. Changing its value yields inequivalent quantum theories. In this paper we will discover a new role of this parameter in the spinfoam dynamics.

A major open theoretical problem is to show that the spinfoam transition amplitudes are peaked on the classical geometries, namely on the solutions of the Einstein equations, in the classical regime. Despite many results are encouraging,\footnote{The classical limit seems to provide the correct low-energy physics in the studied cases \cite{Bianchi:2006uf,Alesci:2007tx,Bianchi:2009ri,Bianchi:2010zs,Bianchi:2011hp,Rovelli:2011kf}.} they are still very partial and stronger tests of the theory are needed. What we would like to have is a concrete spinfoam realization of the formal semiclassical expansion
\begin{align}\label{whatweexpect}
\int DeD\omega\,e^{\frac{i}{\hbar}S(e,\omega)}\sim e^{\frac{i}{\hbar}S(e^0,\omega^0)}
\end{align}
of the path integral in the classical limit $\hbar\rightarrow 0$, where $(e^0,\omega^0)$ is the classical solution determined by the boundary conditions.

Here we propose a tentative mechanism that could solve this problem. Exploiting our suggestion that the Barbero-Immirzi parameter can be used to control the strength of the geometricity constraints,\footnote{The geometricity constraints are the constraints needed in order reduce the spinfoam dynamical variables to the configurations compatible with the metric geometry of the triangulation.} we consider the spinfoam amplitude in the combined flipped limit\footnote{The limit is named after the `flipped' model \cite{Engle:2007qf} where the Barbero-Immirzi parameter is exactly zero. However, we stress that in our case we do not put $\gamma=0$, but expand the amplitude at small $\gamma$.} $\gamma\rightarrow 0$ of small Barbero-Immirzi parameter and classical limit of large areas, and provide good evidence that taking these two limits in the appropriate order the spinfoam amplitude behaves like \eqref{whatweexpect} and thus yields the Einstein equations.

The analysis is partly based on the large spin results of references \cite{Conrady:2008mk,Barrett:2009gg,Barrett:2009mw}.

The paper is organized as follows. First, we review the definition of the EPRL-FK spinfoam model and cast it in a form suitable for the analysis. In the two subsequent sections we perform the flipped limit and the classical limit. Then we extend the discussion to the Lorentzian signature and give more details on the boundary state amplitude. We conclude with some outlooks for future developments.

Throughout the paper we work in units $8\pi G=\hbar=c=1$, but will restore some of these constants when needed. In particular, restoring only $\hbar$ the Planck length is $l_P\simeq\sqrt\hbar$.

\section{The spinfoam amplitude}
We consider the EPRL-FK spinfoam amplitude \cite{Engle:2007wy,Freidel:2007py} for a 2-complex $\sigma$ without matter. We first discuss the Euclidean signature and Barbero-Immirzi parameter in the range $0<\gamma<1$, since we will expand around $\gamma=0$. All the arguments can be easily extended to the Lorentzian signature, discussed later.

Moreover, we restrict for simplicity to a 2-complex which is the 2-skeleton of the dual complex of a simplicial triangulation. The 2-complex is thus made of 0-dimensional vertices $v$, dual to the 4-simplices $v^*$ of the triangulation, 1-dimensional edges $e$, dual to the tetrahedra $e^*$, and 2-dimensional faces $f$, dual to the triangles $f^*$. Faces are oriented and bounded by a cycle of edges. Each edge bounding a face has a source vertex $s(ef)$ and a target vertex $t(ef)$, where source/target is relative to the orientation of the face. We define also a sign symbol $\epsilon_{vef}$ which is $1$ if the orientation of the edge induced by $f$ is ingoing at the vertex, $-1$ otherwise. A face $f$ can be subdivided into portions called wedges, labeled by a vertex-face couple $vf$. If $\epsilon_{vef}$ is 1 we call $e=s(vf)$ the source edge of the wedge $vf$; target edge $e=t(vf)$ otherwise.

We stress that at this stage the triangulation and its dual 2-complex are just combinatorial objects and there is no notion of metric or connection. In particular, the simplices (4-simplices, tetrahedra, triangles) are also combinatorial. The geometry of the triangulation is dynamical and defined by the spinfoam variables labeling the 2-complex. Whenever a geometry is assigned to a simplex we will use the term \emph{geometric} simplex to distinguish it from the purely combinatorial one.

Let us introduce the dynamical variables labeling the 2-complex. For each face $f$ there is an associated half-integer spin $j_f$.  The spin is interpreted as the quantum number of the area operator of the triangle $f^*$. The local gauge group of the Euclidean theory is $SO(4)$, or better its double cover $Spin(4)$ since we are doing quantum mechanics. Thus to each edge-vertex couple we associate a $Spin(4)$ group variable $g_{ev}$. The inverse of these variables will be denoted with $g_{ve}=g_{ev}^{-1}$. More geometrically, the group variables $g_{ev}$ define a connection over the 2-complex. In fact, the combination $g_{v'e}g_{ev}$ can be interpreted as the holonomy of the connection $\omega$ along the path $e$ from the vertex $v$ to the vertex $v'$. It connects the local frame at $v$ to the local frame at $v'$.\footnote{More precisely, since $g_{ev}$ belongs to the spin group, it should be thought as a spin connection. The spin connection is the lift to the spin bundle of the connection $\omega$ defined on the frame bundle.} To each edge-face couple we associate a unit vector $\vec n_{ef}$ in $\mbb R^3$, interpreted as a three-dimensional normal to the triangle $f^*$.

The elementary face amplitude of the EPRL-FK model is defined as
\begin{align}\label{Pf}
P_f=\text{tr}\,\overleftarrow\prod_{e\in f} P_{ef},
\end{align}
where $\overleftarrow\Pi$ is the product (composition) of operators, ordered according to the orientation of the face, and the edge-face operator $P_{ef}$ is given in the usual bra/ket notation by
\begin{align}\label{Pef}
P_{ef}= g_{t(ef),e}Y\ket{j_f,\vec n_{ef}}\bra{j_f,\vec n_{ef}}Y^\dagger g_{e,s(ef)}.
\end{align}
Here $\ket{j,\vec n}$ is the spin-$j$ $SU(2)$ Bloch coherent state \cite{Bloch:1946zza} for angular momentum\footnote{Bloch coherent states were introduced in spinfoams by Livine and Speziale \cite{Livine:2007vk}.} along the direction of $\vec n$. Notice that there is a $U(1)$ phase ambiguity in the notation $\ket{j,\vec n}$, corresponding to a rotation about the axis of $\vec n$. However, this arbitrary phase cancels by multiplication of $\ket{j,\vec n}$ with its conjugate state $\bra{j,\vec n}$. Therefore the edge-face operator \eqref{Pef} is well-defined.

The map $Y$ is the most important ingredient. It is defined as follows. The group $Spin(4)$ has the selfdual/anti-selfdual decomposition $Spin(4)\simeq SU(2)\times SU(2)$. Thus its irreducible representations are labeled by two $SU(2)$ spins $(j^+,j^-)$. The $(j^+,j^-)$ representation is $Spin(4)$-irreducible but is $SU(2)$-reducible, $SU(2)$ acting as a $Spin(4)$ rotation that leaves a reference time 4-vector invariant. We will say more about this reference vector later in this section. Consider the decomposition of a $(j^+,j^-)$ representation into $SU(2)$ irreducibles,
\begin{align}\label{spin4decomposition}
\mathcal H^{(j^+,j^-)}=\bigoplus_{k=|j^+-j^-|}^{j^++j^-}\mathcal H^{(j^+,j^-)}_k,
\end{align}
where the spin $k$ labels the $SU(2)$-irreducible subspaces. Moreover, let the $(j^+,j^-)$ labels be constrained by the equation
\begin{align}\label{ratio}
j^\pm=\frac{1\pm\gamma}{2}j,
\end{align}
where $j$ is a $SU(2)$ spin.\footnote{Because of equation \eqref{ratio} $\gamma$ must be rational and not all spin values are allowed. These restrictions could lead to difficulties for the formulation of the flipped limit studied here. Nevertheless, in the physical model with Lorentzian signature these restrictions are not present and $\gamma$ is any non-zero real number.} The map $Y$ is then defined as the isometric embedding of the spin-$j$ irreducible into the highest weight of the decomposition \eqref{spin4decomposition}, namely into the subspace labeled by $k=j^++j^-=j$.

In fact, the map $Y$, once written in the canonical basis, is nothing else than the Clebsch-Gordan coefficients:
\begin{align}
\bra{j^+j^-,m^+m^-}Y\ket{j,m}=C(j^+j^-j,m^+m^-m),
\end{align}
with $j^\pm$ as in \eqref{ratio}.

As usual in quantum mechanics, the quantum amplitude is the sum over all histories. Thus taking the product of all face amplitudes and summing over the possible labels we get to the spinfoam amplitude
\begin{align}\label{amplitudejn}
W=\sum_{\{j_f\}}\int \!dg_{ev}\!\int \!d\vec n_{ef}\prod_f  d_{j_f}P_f,
\end{align}
where $d_{j_f}=2j_f+1$ is the dimension of the spin-$j$ representation.\footnote{See \cite{Bianchi:2010fj} for a motivation of the choice of the measure factor $d_{j_f}$.} Equivalently, we have the exponential form
\begin{align}\label{amplitudejnS}
W=\sum_{\{j_f\}}\int \!dg_{ev}\!\int \!d\vec n_{ef} \,\Omega\,e^{S},
\end{align}
which resembles a typical path integral, with a complex action given by
\begin{align}\label{actionface}
S=\sum_f S_f=\sum_f \ln P_f
\end{align}
and measure $\Omega=\Pi_f d_{j_f}$. This is the quantum amplitude that defines completely the spinfoam model. 

We can rewrite the action $S$ in a form more convenient for the following analysis. Using the selfdual/anti-selfdual decomposition of the group variables $g=(g^+,g^-)\in SU(2)\times SU(2)$ and the factorization properties of $SU(2)$ coherent states \cite{Perelomov:1986tf} we write
\begin{align}
P_{f}=P^+_{f}\otimes P^-_{f},
\end{align}
where the selfdual/anti-selfdual face amplitudes are
\begin{align}
P^{\pm}_f=\overleftarrow\prod_{e\in f}P^{\pm}_{ef},
\end{align}
with factorized edge-face operators
\begin{align}\label{factorededge}
P^\pm_{ef}=g^\pm_{t(e),e}\ket{\vec n_{ef}}^{\otimes2j^\pm_f}\bra{\vec n_{ef}}^{\otimes2j^\pm_f}g^\pm_{e,s(e)}.
\end{align}
Now the Bloch coherent states are in the fundamental $j=\frac{1}{2}$ representation. We define also the selfdual/anti-selfdual face amplitudes in the fundamental representation as
\begin{align}\label{faceampfund}
\tilde P^\pm_f=\left.P^\pm_f\right|_{j_f=\frac{1}{2}}.
\end{align}
Then the action \eqref{actionface} can be rewritten as
\begin{align}\label{actionsuitable}
S=\sum_f\gamma j_f (\ln \tilde P^+_f-\ln \tilde P^-_f)+j_f (\ln \tilde P^+_f+\ln \tilde P^-_f).
\end{align}
This action will be the starting point for the analysis of the next section.

Notice that in order to write the selfdual/anti-selfdual decomposition and thus define the model (in particular the map $Y$) we implicitly needed a fiducial orthonormal basis formed by three `space' vectors $V^{(1)},V^{(2)},V^{(3)}$ and a `time' vector $N$, together with an orientation. We can take this basis to be the standard one of $\mbb R^4$, together with the standard orientation. The role of the fiducial basis can be viewed in the bivector geometry picture. A bivector $B\in\Lambda^2\mbb R^4$ can be decomposed into two three-dimensional vectors as
\begin{align}
\label{bivectorsplit1}
&B_+^i=(B+B^*)^{IJ}V^{(i)I}N^J,\\
\label{bivectorsplit2}
&B_-^i=(B-B^*)^{IJ}V^{(i)I}N^J.
\end{align}
Observe that the orientation is needed to define the fiducial Hodge $*$ operator. Incidentally, we will use both notations $B^*\equiv *B$ for the Hodge duality in this paper.

As a Lie algebra, the space of bivectors $\Lambda^2\mbb R^4$ is isomorphic to the algebra of rotations $so(4)$.\footnote{Bivectors can be viewed as $4\times4$ antisymmetric matrices, forming a $so(4)$ Lie algebra with product given by the matrix commutator. Exponentiating a bivector yields a $SO(4)$ rotation. In the Lorentzian case it yields a $SO(1,3)$ Lorentz transformation.} So we can use the previous bivector decomposition to define concretely the split $g_{ev}=(g_{ev}^+,g_{ev}^-)$ and the representation-theoretic map $Y$ accordingly. Moreover, we can interpret the couple $(\vec n_{ef},\vec n_{ef})$ as the selfdual/anti-selfdual components of a bivector orthogonal to the fiducial time direction $N$. This bivector is associated to the middle of the edge $e$, and is parallel-transported to the vertices as $g_{ve}(\vec n_{ef},\vec n_{ef})=(g_{ve}^+\vec n_{ef},g_{ve}^-\vec n_{ef})$. We will say more about these bivectors in the section dedicated to the geometrical interpretation.

We have provided the definition of the EPRL-FK model for $0<\gamma<1$ on an arbitrary triangulation, in a form suitable for the purposes of the following analysis.
\section{The flipped\texorpdfstring{ $\gamma\rightarrow 0$}{} limit}
The continuum limit of the theory is defined as the infinite refinement of the triangulation \cite{Rovelli:2010qx}. Here we do not study such a limit but work with a truncation of the theory down to a \emph{finite} number of degrees of freedom, namely with an approximation of the full theory on a fixed and finite triangulation. The spinfoam amplitudes defined on this truncation can be viewed as effective amplitudes obtained by coarse-graining \cite{Markopoulou:2002ja,Bahr:2010cq}.

Given that the truncated amplitudes are not fundamental, there is no reason to believe that they are correct in all physical regimes and a tuning of the parameters of the theory may be needed. In this paper we imagine that the `bare' value of the Barbero-Immirzi parameter is of order 1, and that the renormalized one is close to zero. The possibility of a renormalization of the Barbero-Immirzi parameter has been recently studied in (not so) different contexts \cite{Daum:2010qt,Benedetti:2011nd}.

Therefore we study a regime that we call flipped limit defined as follows:
\begin{align}\label{limitconsidered}
j\rightarrow\infty,\;\gamma\rightarrow 0
\end{align}
with $\gamma j$ kept constant. Notice that according to the area spectrum of loop quantum gravity the quantity $a=\gamma j$ is the physical area in Planck units.\footnote{The area spectrum is rather $a=\gamma\sqrt{j(j+1)}$ but here this difference is irrelevant since we are interested in the large spin regime. We stress also that the flipped limit is taken with $G$ and $\hbar$ constant.} In the regime we are studying this area defines the typical scale of each 4-simplex in the triangulation, and since we work on a finite triangulation with few 4-simplices the scale $a$ must be large in order to describe a macroscopic region of spacetime. The limit of large area is related to the classical limit, discussed in the next section. For the purposes of this section we just imagine that the area is large but kept fixed while performing the flipped expansion. Thus the flipped limit is not a semiclassical expansion but has a different nature, not yet clear to the authors, apart from being possibly related to coarse-graining.

Remember also that the Barbero-Immirzi parameter controls the spacetime discreteness. In particular it controls the area gap and the spacing between the area eigenvalues, thus the limit \eqref{limitconsidered} is a peculiar way of taking the continuous spectrum limit of the area operator.

In order to see the consequences of being in the regime \eqref{limitconsidered} let us rewrite the action \eqref{actionsuitable} in terms of the area variables $a_f=\gamma j_f$ as
\begin{align}\label{actionf}
S=\sum_f a_f (\ln \tilde P^+_f-\ln \tilde P^-_f)+\frac{1}{\gamma}a_f (\ln \tilde P^+_f+\ln \tilde P^-_f)
\end{align}
and notice that this action has the form
\begin{equation}\label{fullaction}
S=S^0+\frac{1}{\gamma}S',
\end{equation}
where $S^0$ and $S'$ do not depend explicitly on $\gamma$. Accordingly, the path integral is rewritten as
\begin{align}\label{amplitudeSplitted}
W=\sum_{\{a_f\}}\int \!dg_{ev}\!\int \!d\vec n_{ef} \,\Omega\,e^{S^0+\frac{1}{\gamma}S'},
\end{align}
where now the action is viewed as a function of the areas, and therefore the sum is over $\gamma$-multiples of the spins.

Now we make an observation that was in fact the main motivation for the development of this work. In the flipped regime the second part $S'$ of the action is multiplied by a large factor $1/\gamma$. Therefore the path integral should be dominated by the stationary phase configurations of $S'$. Intuitively, the second term $S'$ is akin to the Holst term of the action \eqref{Sintro}. 

Thus we expect that the equations of motion arising from the sole term $S'$ contain in particular the simplicial version of the Cartan equation \eqref{Cartan}, that is the vanishing of torsion. As we shall see, this is what happens. More precisely, the term $S'$ yields all the geometricity constraints that reduce the dynamical variables to the metric variables of Regge calculus. These geometricity constraints imply in particular that the on-shell spinfoam connection is torsion-free.

This will be clearer after the study of the equations of motion and their geometrical interpretation. So let us start our analysis of the flipped limit in the Euclidean signature, taking our main source of inspiration from the references \cite{Conrady:2008mk,Barrett:2009gg}. Let us stress that in this paper we restrict our study to the nondegenerate solutions. Therefore, as in \cite{Conrady:2008mk}, we will disregard the solutions that can be interpreted as degenerate metrics of the triangulation.

As we said, we are interested in the asymptotic behavior of the spinfoam amplitude \eqref{amplitudeSplitted} in the limit $\gamma\rightarrow 0$ with physical areas kept constant. Here we need to clarify what we mean by `constant'. Obviously, the area variables $a_f$ are not free parameters because we are summing over them, so we cannot really keep them constant. What we keep constant is the average value of each $a_f$, which is supposed to be of the order of the boundary areas. In fact the boundary state, which is a free input of the amplitude, can be used to fix a scale $a$ on the boundary of the triangulation (see the dedicated section). This scale is the typical area of the triangles on the boundary. Then we make the hypothesis that the internal areas $a_f$ are all peaked on values of the order of $a$, $a_f\sim a$, namely that the internal areas of the order of $a$ are the ones contributing the most to the spinfoam amplitude in the regime studied. This is the main working hypothesis of the paper.

A second working hypothesis is that we can take the variations of $S'$ with respect to the quantized areas $a_f$. In other words we need to approximate the sum over $a_f$ with an integral. This can be done only if the fluctuations of the action around the stationary points are essentially within a region that contains a sufficiently large number of area eigenvalues. Let us suppose we can make this approximation.

Having this in mind, we proceed with the stationary phase evaluation of the spinfoam amplitude \eqref{amplitudeSplitted} in the flipped regime. The action term $S'$ is complex and we need a generalization of the stationary phase method.\footnote{In \cite{Barrett:2009gg} a similar method is used, except that the area variables are not taken into account in the stationary phase, because there the analysis is done for a single 4-simplex at fixed areas.} In particular the real part of $S'$ is nonpositive so we require that a stationary point maximizes the real part of $S'$, otherwise the amplitude would be exponentially suppressed for small $\gamma$.

We call critical points the stationary points satisfying also the real part condition
\begin{align}\label{Re0}
\text{Re}\,S'=0.
\end{align}
Let us write this condition in terms of the spinfoam configurations $(a_f,\vec n_{ef},g_{ev})$. Both $P_f^+$ and $P_f^-$ have nonpositive real part, and the condition \eqref{Re0} holds when $\text{Re}\,P_f^+=\text{Re}\,P_f^-=0$ for all faces. Using that the scalar product of Bloch coherent states satisfies $|\langle \vec n_1|\vec n_2\rangle|=\frac{1+\vec n_1\cdot\vec n_2}{2}$ we get easily that \eqref{Re0} is equivalent to
\begin{align}
\label{critical}g^\pm_{ve}\vec n_{ef}=g^\pm_{ve'}\vec n_{e'f},
\end{align}
where $e=s(vf)$ and $e'=t(vf)$ are the two adjacent edges of the face $f$, meeting at the vertex $v$. Here $g_{ve}^\pm$ acts in the vector representation on $\mbb R^3$. Interpreting $g_{ve}=(g_{ve}^+,g_{ve}^-)$ as a transport, namely as a $Spin(4)$ connection, the previous equation is a matching condition under transport. Namely, the bivector $(\vec n_{ef}, \vec n_{ef})$ in the frame at $e$ and the bivector $(\vec n_{e'f}, \vec n_{e'f})$ in the frame at $e'$ must match when transported to the common frame at $v$. We will say more about this in the following section.

Lifting \eqref{critical} to the corresponding Bloch states, we have the equation
\begin{align}\label{brackets}
g_{e'v}^{\pm}g^\pm_{ve}\ket{\vec n_{ef}}=e^{i\theta^\pm_{vf}}\ket{\vec n_{e'f}},
\end{align}
equivalent to \eqref{critical}.

Now let us impose the stationary phase conditions. Varying $S'$ with respect to $g_{ev}$ and evaluating on a solution of \eqref{brackets} we get
\begin{align}\label{closure}
\left.\delta_{g_{ev}} S'\right|_{crit.}=0\Rightarrow\sum_{f\in e}\epsilon_{vef}a_{f}\vec n_{ef}=0,
\end{align}
which expresses the closure relation for the tetrahedron dual to the edge $e$. In fact, at each edge $e$, the closure of the four vectors $\epsilon_{vef}a_f\vec n_{ef}$ allows us to interpret them as the normals, all outward or all inward, to the four triangular faces of a geometric tetrahedron in $\mbb R^3$, where the areas of the triangles are given by the norm $\|\epsilon_{vef}a_f\vec n_{ef}\|=a_f$.

Finally, varying with respect to the unit vectors $\vec n_{ef}$ does not give further information, because it leads to an empty equation. The variation of the areas $a_f$ will be considered later, after the following section.
\section{Geometrical interpretation}
Before completing the flipped limit expansion, let us open a discussion on the geometrical interpretation of the equations of motion found so far. A critical point can be interpreted as the geometry of a four-dimensional Regge triangulation\footnote{A Regge triangulation is a simplicial complex with a continuous piecewise flat metric such that the 4-cells are isometric to geometric 4-simplices in $\mbb R^4$. Curvature is distributional and concentrated on the triangles, the `hinges' of the tringulation.} using the following construction.

Given a critical configuration $(a_f, \vec n_{ef}, g^\pm_{ev})$, let us interpret the three-dimensional vectors
\begin{align}\label{Bpm}
\vec A^\pm_{vf}=a_f g^\pm_{ve}\vec n_{ef}
\end{align}
as the selfdual ($+$) and anti-selfdual ($-$) components of a four-dimensional bivector $A_{vf}=(\vec A^+_{vf},\vec A^-_{vf})$. As explained previously, in order to define this decomposition we have used a fiducial oriented basis, the standard one of $\mbb R^4$. Let us also define the dual bivector $B_{vf}=A^*_{vf}$ using the fiducial Hodge operator. 

We want to interpret $A_{vf}$ as the simplicial analogue of the tetradic quantity $e\wedge e$ in the Holst action \eqref{Sintro}. In order to see why this interpretation is correct let us concentrate for a moment on a single vertex $v$ and recall some results from \cite{Barrett:2009gg}.

First, notice that thanks to the matching condition \eqref{critical} we have dropped the edge label $e$ from the definition of the bivector. Thus for each vertex we have ten bivectors $A_{vf}$ and not twenty.

Second, the bivectors $A_{vf}$ are simple and cross-simple, by construction. They are clearly simple\footnote{A bivector $B\in\Lambda^2\mbb R^4$ is called a simple bivector if it is the wedge product $B=X\wedge Y$ of two 4-vectors $X$ and $Y$. Equivalently, if $B^{IJ}(B^*)_{IJ}=0$. Therefore a simple bivector has an associated 2-plane, the plane determined by $X$ and $Y$.} because $\|A_{vf}^+\|=\|A_{vf}^-\|$ holds. Thus $A_{vf}$ determines a two-dimensional plane. Cross-simplicity\footnote{Consider a collection of bivectors $B^{(i)}$ $(i=1,2,\ldots)$. These are said cross-simple if any combination of two of them is simple, that is if $B^{(i)}+B^{(j)}$ is simple for all $i\neq j$.} holds at each edge because the four bivectors $A_{vf}$ with $f\ni e$ share the same orthogonal direction, the direction $g_{ve}N$, where we remember that $N=(1,0,0,0)$ is the reference time direction of the fiducial basis. This orthogonality property can be seen by going back to the time gauge: the four bivectors $g_{ev}A_{vf}$ with $f\ni e$ are orthogonal to the reference direction because for each one of them the selfdual and anti-selfdual components are the same.

Third, from the closure critical equation \eqref{closure} the area bivectors close at each edge: $\sum_{f\ni e}\epsilon_{vef}A_{vf}=0$.

The Barrett-Crane reconstruction theorem \cite{Barrett:2009gg} states that given a set of ten bivectors which are simple, cross-simple, nondegenerate,\footnote{In this paper we do not study the degenerate sector, namely we disregard the critical points that determine degenerate 4-simplices.} and satisfy the closure condition, there is a unique oriented geometric 4-simplex in $\mbb R^4$, up to translation and spacetime inversion, such that its area bivectors coincide with the given ones.

A geometric 4-simplex has five external unit normals $N_e$, the four-dimensional normals to the five tetrahedra of its boundary, pointing towards the exterior. The area bivectors of a geometric 4-simplex are then defined in terms of its unit normals $N_e$ as
\begin{align}\label{areabivfromnormals}
A_{vf}=a_f *_{\epsilon_v}\frac{N_{s(vf)}\wedge N_{t(vf)}}{\|N_{s(vf)}\wedge N_{t(vf)}\|},
\end{align}
where $a_f $ is the area of the triangle, $\epsilon_v=\pm 1$ parametrizes the orientation of the reconstructed 4-simplex (plus for the fiducial orientation, minus for the opposite one) and $*_{\epsilon_v}:=\epsilon_v *$ is the Hodge operator with orientation $\epsilon_v$.

Thus to a nondegenerate critical configuration we can associate at each vertex a (almost) unique geometric 4-simplex. From this theorem it follows that our on-shell bivectors $A_{vf}$ are in fact the area bivectors of a geometric 4-simplex and thus can be written as \eqref{areabivfromnormals} or equivalently as
\begin{align}\label{areabiv}
A_{vf}=\frac{1}{2}E_{vb}\wedge E_{vb'},
\end{align}
where $E_{vb}$ and $E_{vb'}$ are two side vectors ($b$ labels the side) of the triangle $f^*$.\footnote{In formula \eqref{areabiv} we can choose any two sides $b$ and $b'$ out of the three sides of the triangle $f^*$, but the orientation of the two side vectors must be chosen so as to be consistent with formula \eqref{areabivfromnormals}.} Notice that the area bivector $A_{vf}$ is normalized to the area of the triangle,\footnote{In this paper, the norm of a bivector $B$ is defined by $\|B\|^2=\frac{1}{2}B^{IJ}B_{IJ}$.} lies in the plane of the triangle and its orientation is the one determined by the 4-simplex orientation $\epsilon_v$.\footnote{The orientation of the area bivector $A_{vf}$ (the orientation of its plane) is defined as the one pull-backed from the orientation of the source tetrahedron $s(vf)^*$ (the orientation of its three-dimensional plane). In turn the orientation of the tetrahedron is the one pull-backed from the orientation of the 4-simplex (the orientation $\epsilon_v$ of $\mbb R^4$ determined by the Barrett-Crane theorem).}

The next step is the reconstruction of the full Regge triangulation from the individual 4-simplices, along with the full cotetrad and on-shell connection. This is the simple though original part of the reconstruction (but notice that a similar analysis was already done in \cite{Conrady:2008mk}).

Two neighboring 4-simplices of a Regge triangulation share a tetrahedron. Let us describe the Regge triangulation with local Cartesian charts, each covering a single 4-simplex. When viewed in different charts, the common tetrahedron is described in general by two different tetrahedra in $\mbb R^4$, but their shape is the same. Namely, they are connected by a Poincar\'e transformation which defines the transition function between the two charts. 

In principle, in spinfoams this property may not hold and the two reconstructed tetrahedra could have a different shape. However this is not the case for the models considered in this paper, because the three-dimensional geometry of a shared tetrahedron $e^*$ is described by a common set of four closing normals $\epsilon_{vef}a_f\vec n_{ef}=-\epsilon_{v'ef}a_f\vec n_{ef}$. Thus for each edge $e$ (bounded by $v$ and $v'$) there is a $O(4)$ rotation $U_{v'v}$ that brings the tetrahedron $e^*$ of the geometric 4-simplex $v^*$ to the tetrahedron $e^*$ of the geometric 4-simplex ${v'}^*$ (up to translation), and their four-dimensional outward normals in the anti-parallel configuration. Therefore the individual reconstructed 4-simplices are in fact the 4-cells of a Regge triangulation, viewed in local Cartesian charts.

This means that the side vectors $E_{vb}$ in \eqref{areabiv} are a consistent simplicial cotetrad for all the triangulation and $U_{v'v}$ is the Levi-Civita transport, namely the unique torsion-free connection.

In order to find the precise relation between the Levi-Civita connection and the $Spin(4)$ connection variables $g_{ev}$ we first notice that the reconstructed 4-simplices can have different orientations \cite{Hellmannprivate}. This is because from any solution $(a_f,\vec n_{ef},g_{ev})$ of the critical equations we can generate another solution by a local parity transformation at an arbitrary vertex $v$ (see \cite{Barrett:2009gg}). The local parity at a vertex $v$ corresponds to switching simultaneously the selfdual and anti-selfdual parts of the five group variables $g_{ev}$. The oriented geometric 4-simplex reconstructed from this parity-related solution is related to the original one by spatial inversion $P:\vec x\rightarrow-\vec x$ and a flip of the orientation $\epsilon_v\rightarrow -\epsilon_v$.

We can apply this elementary local transformation to an arbitrary subset of vertices. Thus $(a_f,\vec n_{ef},g_{ev})$ labels a (finite) class of $2^V$ parity-related solutions, where $V$ is the number of vertices.

Now we have that two neighboring reconstructed 4-simplices can be connected via a $SO(4)$ rotation only if their orientations match. When this holds for all the edges, namely when all geometric 4-simplices have the same orientation, the reconstructed cotetrad $E_{vb}$ is globally oriented and the evaluation of the spinfoam action on this solution yields the proper Regge action as we shall see in a moment, up to a global sign. For all the other solutions in the parity-related class, the simplicial cotetrad flips its orientation somewhere in the triangulation and we have a generalized Regge action.\footnote{The fact that the 4-simplex orientation can flip across the triangulation is to be expected since the model is based on the action \eqref{Sintro} without restrictions on the orientation of the cotetrad. Thus orientation flip is not strange and is not related to the quantum theory itself.}

Thus the $O(4)$ Levi-Civita connection in terms of the spinfoam group variables can be easily shown to be
\begin{align}\label{LCwithSFgroup1}
U_{v'v}=\epsilon_e U(g_{v'e}g_{ev})
\end{align}
for matching orientations, and
\begin{align}\label{LCwithSFgroup2}
U_{v'v}=\epsilon_e U(g_{v'e})PU(g_{ev})
\end{align}
for non-matching orientations, where the map $U$ on the right hand side is the covering map, $P$ is the spatial inversion, and the sign $\epsilon_e$ represents a spacetime inversion ambiguity $PT: x^\mu\rightarrow-x^\mu$.

It is not difficult to understand the previous relations \eqref{LCwithSFgroup1} and \eqref{LCwithSFgroup2}. In fact, since the spinfoam group variables $g_{ev}$ belong to $Spin(4)$, the covering of $SO(4)$, the holonomy $g_{v'e}g_{ev}$ must be orientation-preserving. However, the Levi-Civita holonomy $U_{v'v}$ is in general in $O(4)$, and this is why we need a parity insertion in the case \eqref{LCwithSFgroup2} of non-matching orientations.

Concerning the inversion sign, the following equation holds:
\begin{align}
\epsilon_e=-\epsilon_{ve}\epsilon_{v'e},
\end{align}
where $\epsilon_{ve}=\pm 1$ parametrizes the time orientation (plus for future and minus for past) of the outward normal to the tetrahedron $e^*$ of the reconstructed 4-simplex. This simple expression of the sign $\epsilon_e$ as well as the formulas \eqref{LCwithSFgroup1} and \eqref{LCwithSFgroup2} are derived in \cite{Perini:2012nd}.

Observe that even in this Euclidean context we can conveniently though improperly talk about time orientation of the 4-simplex normals $N_e$, in the following sense: since we must have $g_{ve}N_e=\pm(1,0,0,0)$, where $g_{ve}$ is the critical group element, we say that $N_e$ is future or past oriented depending on this sign. Moreover, observe that in the Lorentzian case a wedge $vf$ is called thick wedge if the two outward normals $N_{s(vf)}$ and $N_{t(vf)}$ are both future oriented or both past oriented, thin wedge if they are oppositely oriented, and we shall borrow this language in the Euclidean theory.

Now notice that a single sign $\epsilon_{ve}$ is sent to its opposite under the inversion ambiguity of the reconstruction theorem. However, to each wedge $vf$ of the face we associate the combination 
\begin{align}\label{wedgesign}
\epsilon_{vf}:=-\epsilon_{v,s(vf)}\epsilon_{v,t(vf)},
\end{align}
which is invariant under inversion of the 4-simplex $v^*$. It is positive for a `thin' wedge and negative for a `thick' wedge.

Spacetime curvature is captured by the holonomy of the Levi-Civita connection along the loop that bounds a face, which is the loop that encircles the `hinge' $f^*$. So multiplying all the edge signs of a face we define the face sign as
\begin{align}\label{facesign}
\epsilon_f:=\prod_{e\in f}\epsilon_e=\prod_{v\in f}\epsilon_{vf},
\end{align}
which is unambiguous and well-defined in terms of the critical configuration only, because of the invariance of \eqref{wedgesign}. Notice that $\epsilon_f=1$ if the face has an even number of thick wedges, $-1$ for an odd number. In \cite{Han:2013gna} a geometry is called time-oriented or time-unoriented at the face $f$ depending on the value of $\epsilon_f$. In the Lorentzian signature, this is equivalent to saying that the Levi-Civita holonomy along the loop around the face is an orthochronous Lorentz transformation for $\epsilon_f=1$ (not necessarily a proper one because of the parity insertion!) and a non-orthochronous transformation for $\epsilon_f=-1$.

Let us compute the loop holonomy of the spinfoam group variables on a critical configuration. Using \eqref{brackets} recursively for the cycle of edges around a face we get that the spinfoam loop holonomy $g_{vf}$ has the form
\begin{align}\label{sfloop}
U(g_{vf})=e^{\tilde\Theta_f B_{vf}+\Theta_f^*B^*_{vf}},
\end{align}
where the loop begins at a vertex $v$ and is oriented like the face, $B_{vf}=A_{vf}^*$, and we have defined the angles
\begin{align}
\label{Thetatilde}&\tilde\Theta_f:=\sum_{v\in f}\theta^+_{vf}-\theta^-_{vf},\\
\label{Thetastar}&\Theta^*_f:=\sum_{v\in f}\theta^+_{vf}+\theta^-_{vf}.
\end{align}
Thus $U(g_{vf})$ decomposes into a rotation by $\tilde\Theta$ on the plane orthogonal to the triangle $f^*$ of the geometric 4-simplex $v^*$ and a rotation by $\Theta^*$ on the plane of the triangle, and preserves both $B_{vf}$ and $B^*_{vf}$. Now we recall from \cite{Barrett:2009gg} (Lemma 5) that the phases in \eqref{brackets} are related to the dihedral angles in the following way:
\begin{align}\label{dihedral}
\theta^+_{vf}-\theta^-_{vf}=\epsilon_v\Theta_{vf},
\end{align}
where we remember that $\epsilon_v$ parametrizes the orientation of the reconstructed 4-simplex $v^*$, and $\Theta_{vf}$ is the four-dimensional dihedral angle of $v^*$ between the two tetrahedra sharing the triangle $f^*$.

Since the deficit angle $\Theta_f$ of Regge calculus is just ($2\pi$ minus) the sum of the dihedral angles, we find that $\tilde\Theta_f$ is not in general the deficit angle. It is $\pm\Theta_f$ when all the 4-simplices of the face $f$ are consistently oriented, the minus being for the $\epsilon_v=-1$ orientation, whereas in the case of mixed orientations the interpretation of $\tilde\Theta_f$ as deficit angle is less natural. It is a generalized deficit angle, in the sense of Barrett and Foxon \cite{Barrett:1993db} (see also \cite{Dowdall:2009eg}).

On the other hand the angle $\Theta^*_f$ describes a rotation on the same plane of the triangle $f^*$ and can be interpreted as spacetime torsion. Therefore it must vanish on the solutions of the equations of motion (up to a possible $\pi$ rotation as we shall see) if the theory is to yield general relativity. This can be seen in different ways. For example we can compute $U(g_{vf})$ on the parity-related solution where all 4-simplices of the face have the same orientation and notice that $\Theta^*_f$ is unchanged if we exchange its $+$ and $-$ parts \eqref{Thetastar}, that is $\Theta^*_f$ is parity invariant. Then, given that the Levi-Civita holonomy is torsion-free we conclude that the $Spin(4)$ loop holonomy is torsion-free as well, regardless of the orientations of the reconstructed 4-simplices.

More precisely, consider a solution which is globally oriented over the face. Using \eqref{LCwithSFgroup1} for the loop we have
\begin{align}\label{fullholon}
U_{vf}=\epsilon_f U(g_{vf}),
\end{align}
where the face sign $\epsilon_f$ is defined in \eqref{facesign}. Now since by definition the Levi-Civita loop holonomy preserves the triangle $e^*$ of the geometric 4-simplex $v^*$, it follows that $U(g_{vf})$ does \emph{not} preserve the triangle whenever $\epsilon_f=-1$; in such cases the only way it can act on the plane of the triangle is by a $\pi$ rotation, because in this way the $\pi$ rotation combined with the inversion $\epsilon_f$ gives back the triangle-preserving Levi-Civita connection.

By the parity-invariance of $\Theta^*_f$ this argument must be valid for general 4-simplex orientations, so we have found that the torsion angle is
\begin{align}\label{Thetastarcrit}
\Theta_f^*=
\begin{cases}
0&\epsilon_f=1\\
\pi&\epsilon_f=-1
\end{cases}
\end{align}
on a critical configuration.\footnote{The fact that $\Theta^*_f$ is non-zero ($\pi$) in some cases does not mean that there is some spacetime torsion, because in such cases the Levi-Civita loop holonomy $U_{vf}$ still preserves the triangle $f^*$, but the spinfoam loop holonomy does not! Therefore the complete torsion is always zero on a critical configuration.}

Finally, defining $\eta_f=0,1$ as $\epsilon_f=e^{i\pi\eta_f}$, the on-shell spinfoam loop holonomy can be written as
\begin{align}\label{sfloopWOtorsion}
U(g_{vf})=e^{\tilde\Theta_f B_{vf}+\eta_f \pi B^*_{vf}},
\end{align}
where $\tilde\Theta_f$ is the generalized deficit angle.
\section{Consequences of the flipped limit}
Now we come back to the the small Barbero-Immirzi parameter expansion. So far we have derived all the critical equations except one. Indeed, we are left with the variations with respect to the areas $a_f$. Notice that this is the problematic point raised in the more common large area expansions found in the literature (see e.g. \cite{Bonzom:2009hw, Hellmann:2013gva, Perini:2012nd} but also \cite{Mamone:2009pw}). If the areas go large regardless of the value of $\gamma$ the requirement of stationary phase with respect to $a_f$ would be satisfied only for vanishing generalized deficit angles\footnote{See \cite{Barrett:1993db} for a geometrical interpretation of vanishing $\tilde\Theta_f$ in the case of mixed orientations.} $\tilde\Theta_f=0$, which is in contradiction with the correct classical limit: curved geometries are allowed in general relativity even if in the considered region there is no matter.

Instead here we are studying the flipped limit of small Barbero-Immirzi parameter where the areas are \emph{not} scaled. In our case the relevant part of the action is $S'$. Varying $S'$ with respect to $a_f$ we get
\begin{align}\label{areavariation}
\left.\frac{\partial S'}{\partial a_f}\right|_{crit.}=\frac{i}{\gamma}\Theta^*_f=0,
\end{align}
where we have used \eqref{brackets} and \eqref{Thetastar}. Using \eqref{Thetastarcrit}, this stationary phase condition is automatically satisfied for all faces if $\epsilon_f=1$ everywhere, namely if the reconstructed triangulation is globally time-oriented. The time-unoriented configurations are then suppressed in the flipped limit. (We stress that the suppression of the time-unoriented solutions was argued for the first time by M. Han in a recent analysis \cite{Han:2013ina} and was not noticed in the previous versions of this paper)

As a remark, the reader should not confuse the time-orientedness $\epsilon_f$ of the faces with the spacetime orientation $\epsilon_v$ of the 4-simplices. The previous equations do not imply the suppression of the non-matching 4-simplex orientations.

Now we evaluate the full action at a (time-oriented) critical configuration. Putting all together, using again \eqref{brackets} and \eqref{Thetatilde}, we have that the full action $S$ at a critical point of $S'$ yields
\begin{align}
S|_{crit.}=i S_R=i \sum_f a_f \tilde\Theta_f,
\label{Sfonshell}
\end{align}
namely the Regge action $S_R$ of general relativity \cite{Regge:1961px,Williams:1991cd}. More precisely, it is a generalized Regge action because the angle $\tilde\Theta$ is a generalization of the standard deficit angle $\Theta$ of Regge calculus. This generalization takes into account the orientation of the 4-simplices. The generalized deficit angle $\tilde\Theta_f$ encodes the spacetime curvature distributed on the triangle $f^*$. Its relation with the spinfoam loop holonomy around the face is given by formula \eqref{sfloopWOtorsion}.

We observe that the appearance of the generalized deficit angle is very natural here because the classical action on which the EPRL-FK spinfoam model is based is the tetradic action \eqref{Sintro}, not the metric Einstein-Hilbert action, and tetrads can have both orientations (with respect to a fiducial one) unless a restriction is imposed by hand in the path integral.

The exponential of $i$ times the generalized Regge action $S_R$ is the contribution of a single critical point to the asymptotics of the spinfoam amplitude in the flipped limit. Now the stationary phase method requires to sum over all critical points. In general this sum is an integral over the critical surface, defined in the following way. Suppose we are given a multidimensional integral of the form
\begin{align}
I=\int \!dx \,a(x) e^{\lambda f(x)},
\end{align}
and we want to compute its asymptotic expansion for large positive $\lambda$. Suppose the action $f$ has nonpositive real part and a surface $\mathcal C$ of critical points $y\in\mathcal C$, with $\text{Re} \,f(y)=0$ and $\partial f/\partial x^i|_y=0$. Then we have the following expansion:
\begin{align}\label{generalasym}
I\sim e^{\lambda f(y_0)}\int_\mathcal C d\mu(y) \frac{a(y)}{\sqrt{\text{det} H^{\bot}(y)}}(1+\mathcal O(1/\lambda)),
\end{align}
where $H^\bot$ is the Hessian matrix of $f$ restricted to the directions orthogonal to the critical surface with respect to some metric, and $\mu$ is the measure induced on the critical surface by the same metric. In the factor on the left of the previous integral the action is evaluated on a  reference critical point $y_0$ (but of course it is independent of the choice).

Using the general formula \eqref{generalasym} of the stationary phase method, with $\lambda=1/\gamma$, $f=S'$, and $a=\exp(S^0)$ we get
\begin{align}\label{previousintegral}
W= \int_{\mathcal C} d\mu(a_f,\vec n_{ef},g_{ev}) \,e^{i S_R}(1+\mathcal O(\gamma)),
\end{align}
where $\mathcal O(\gamma)$ denotes the corrections containing the next-to-leading orders of the asymptotic approximation\footnote{The corrections in $\gamma$ should contain also the error we made by approximating the sum over spins with an integral. However, this error is expected to decay at least exponentially for $\gamma\rightarrow 0$.} and all the spinfoam path integral measure factors have been reabsorbed into $\mu$. Observe that the integration region $\mathcal C$ contains the contributions from all the possible joint orientations $\epsilon=(\epsilon_{v_1},\epsilon_{v_2},\ldots)$ of the 4-simplices.

This is our first result. The spinfoam amplitude in the flipped limit regime yields a path integral weighted with the exponential of the generalized Regge action, up to $\gamma$-corrections.

This result resonates with similar findings in the context of the computation of the spinfoam graviton propagator. In particular, as first shown in \cite{Bianchi:2009ri}, for small $\hbar$ (or better large areas) the spinfoam 2-point function $G(x,y)$ of the Euclidean EPRL-FK model has the form
\begin{align}\label{propagator}
G(x,y)= \frac{R+\gamma X+\gamma^2 Y}{|x-y|^2}+\hbar\text{-corr.},
\end{align}
where $R$ is the matrix of area-angle correlations of Regge calculus, and the other two terms $X$ and $Y$ are the leading order $\gamma$-corrections. Therefore in the flipped limit $\gamma\rightarrow 0$ the spinfoam graviton propagator matches the one of Regge quantum gravity and, most interestingly, the one of standard perturbative quantum field theory of gravitons. In the light of the present analysis, the result \eqref{propagator} is more understandable.

It is important to observe that on the critical configurations one can use length variables in place of the area variables. As we have seen before a critical configuration\footnote{Actually, a class of parity-related critical configurations.} $(a_f,\vec n_{ef},g_{ev})$ is naturally interpreted as a Regge triangulation, hence there must exist an assignment of lengths $l_b$ for the sides $b$ of the triangles such that the areas $a_f$ coincide with the areas computed from the lengths $l_b$,
\begin{align}
a_f=a_f(l_b),
\end{align}
for example via Heron's formula. In other words the critical equations select only the Regge-like area configurations.\footnote{It is well-known that a generic assignment of areas $a_f$ to the faces of the 2-complex $\sigma$ may not be consistent with the geometry of a Regge triangulation \cite{Barrett:1997tx,Makela:2000ej,Dittrich:2008va}. In fact a Regge triangulation is completely described by its lengths, and in general there may not exist an assignment of lengths $l_b$ such that $a_f=a_f(l_b)$.}

Exploiting this, we can parametrize the critical surface $\mathcal{C}$ with the side lengths $l_b$ and the joint orientations $\epsilon=(\epsilon_{v_1},\epsilon_{v_2},\ldots)$ of the 4-simplices. Restoring also the $\hbar$ dependence, we rewrite \eqref{previousintegral} as
\begin{align}\label{intlengths}
W= \sum_{\epsilon}\int \!d\tilde\mu(l_b) \,e^{\frac{i}{\hbar} S^{\epsilon}_R(l_b)}(1+\mathcal O(\gamma\hbar)),
\end{align}
where we have written the generalized Regge action as an explicit function of the lengths,
\begin{align}\label{Reggelengths}
S^{\epsilon}_{R}(l_b)=\sum_f a_f(l_b)\tilde\Theta^{\epsilon}_f(l_b),
\end{align}
similarly to the original formulation of Regge \cite{Regge:1961px}. We have also made the dependence on the joint orientations $\epsilon$ explicit.

We have found that in the regime of small Barbero-Immirzi parameter the spinfoam theory on a finite triangulation approaches a theory which resembles the standard quantization of Regge gravity, with a specific path integral measure. In this regime the effective degrees of freedom are the lengths (see \cite{Rocek:1982fr} for an introduction to quantum Regge gravity).

This result is not completely unexpected. Using general considerations on the Holst action \eqref{Sintro} one could already argue that since for small $\gamma$ the Holst term becomes large then the path integral should be dominated by the solutions of the equations of motion arising by varying this term (akin to our term $S'$ in the spinfoam action). It is well known in classical general relativity that the Cartan equation \eqref{Cartan} can be obtained in this way, varying the sole Holst term with respect to $\omega$. Thus $\gamma$ controls in particular the strength of the torsionless constraint at the quantum level and in the limit $\gamma\rightarrow 0$ the wildly fluctuating connection $\omega$ freezes and becomes the torsion-free one, $\omega=\omega(e)$. In other words, the first-order quantum theory with independent tetrad and connection reduces effectively to the second-order theory with only tetrads. We have found a concrete realization of this phenomenon in spinfoams.

As a next step, we will study the classical regime of large areas, which basically reduces to computing the well-known equations of motion of Regge calculus, and the subtleties of taking the flipped limit and the classical limit simultaneously.
\section{The classical\texorpdfstring{ $\hbar\rightarrow 0$}{} limit}
As a further step, let us study the semiclassical approximation. The formal classical limit $\hbar\rightarrow 0$ can be understood more physically as the limit of large geometries, namely as the regime
\begin{align}
a_f\gg l^2_P,
\end{align}
where all areas $a_f$ have the same large order of magnitude $a_f\sim a$, large with respect to the Planck length $l_P$ squared. We are working with a truncation of the full theory to few 4-simplices and therefore a macroscopic region of spacetime can only be described by large individual areas. This regime can be studied using a stationary phase method for the Regge action \eqref{Reggelengths}, where the areas play the role of the large parameter controlling the rapidity of oscillation.

Now since we are also taking the flipped limit we need to be careful because the two expansions are not independent. In the previous section we sent $\gamma$ to zero while keeping the areas fixed, whereas here we are interested in large areas. How can we reconcile the two regimes? The answer is simple: we need to choose some large but fixed scale for the areas, say $10^6$ in Planck units, so that the semiclassical approximation of quantum Regge gravity will be good for the desired accuracy, and $\gamma$ arbitrarily small so that the approximation in which spinfoams reduce to quantum Regge gravity is justified.\footnote{On the other hand, if we first fixed a small $\gamma$ and then let the areas be arbitrarily large we would be in the usual large spin regime, which suppresses the spinfoam amplitude unless the deficit angles are zero \cite{Bonzom:2009hw}; namely, the spinfoam amplitude would be peaked only on the flat spacetime solutions, no matter how small is $\gamma$, because the large spin asymptotics would dominate.}

An equivalent way of understanding this particular regime is by looking at the form of the action \eqref{fullaction}, which is made of two terms, $S^0$ and $S'/\gamma$. As we already said, loosely speaking they correspond to the Einstein-Hilbert term and the Holst term of the classical action \eqref{Sintro}. Clearly, the flipped expansion is justified, even for large areas, only if the second term oscillates much more rapidly than the first. Therefore, since the large quantities used in the stationary phase expansion are the areas $a_f$ for the Einstein-Hilbert term and the ratio $a_f/\gamma$ for the Holst term, we need to have
\begin{align}\label{compactregime}
l^2_P\ll a_f\ll a_f/\gamma,
\end{align}
which expresses in a compact form the regime of validity of the approximations studied in this paper. This regime can be selected in principle by appropriate boundary conditions and a fine tuning of the Barbero-Immirzi parameter, as explained later.

Notice also that the expansion of the spinfoam amplitude in this regime is to be interpreted more physically as an expansion in powers of $l_P^2/a$, for the classical limit, and in powers of $\gamma l_P^2/a$, for the flipped limit, with $\gamma\ll 1$.

So let us perform the semiclassical expansion of the effective path integral \eqref{intlengths}. By the stationary phase method, applied to each term of the finite sum over the joint orientations $\epsilon$, we select the solutions of the classical equations of motion obtained by varying the generalized Regge action \eqref{Reggelengths} with respect to the lengths.

We observe that the Schl\"afli identity \cite{Regge:1961px}
\begin{align}\label{Schlafli}
\sum_{f\ni v}\ a_f\frac{\partial\Theta_{vf}}{\partial l_b}=0,
\end{align}
where the quantities $\Theta_{vf}$ are dihedral angles, arises even for non-matching 4-simplex orientations. In fact the contribution of the variation of the generalized deficit angles to the variation of the generalized Regge action is
\begin{align}\nonumber
\sum_f a_f\frac{\partial \tilde\Theta_{f}}{\partial l_b}=\sum_v\epsilon_v\sum_{f\ni v}\ a_f\frac{\partial\Theta_{vf}}{\partial l_b},
\end{align}
which vanishes thanks to the Schl\"afli identity \eqref{Schlafli}. Thus the generalized Regge equations take the simple form
\begin{align}\label{Ricciflat}
\sum_{f}\frac{\partial a_f}{\partial l_b}\tilde\Theta^\epsilon_f=0
\end{align}
and determine a relation between the generalized deficit angles of different faces. These are the simplicial version of the Einstein equations \eqref{Einsteintetrad}.

Therefore, provided that the few working hypotheses of this paper are correct, the spinfoam amplitude in the regime \eqref{compactregime} is dominated by configurations that solve the Einstein equations.
\section{Lorentzian signature}
The Lorentzian EPRL spinfoam model \cite{Livine:2007ya} has the same form of the Euclidean one, \emph{mutatis mutandis}, so the analysis is virtually identical to the Euclidean case. The formalism and the strategy is partly based on \cite{Barrett:2009mw}. We provide the basic formulas needed for the Lorentzian signature and the main arguments for the reconstruction of geometry.

As before, the spinfoam amplitude is
\begin{align}
W=\sum_{\{j_f\}}\int \!dg_{ev}\!\int \!d\vec n_{ef}\prod_f  d_{j_f}P_f,
\label{LorZ}
\end{align}
with face amplitude $P_f$ given by the trace of operators
\begin{align}\label{PfLOR}
P_f=\text{tr}\,\overleftarrow\prod_{e\in f} P_{ef},
\end{align}
where the edge-face operators are
\begin{align}\label{PefLOR}
P_{ef}= g_{t(ef),e}Y\ket{j_f,\vec n_{ef}}\bra{j_f,\vec n_{ef}}Y^\dagger g_{e,s(ef)}.
\end{align}

The gauge group is now the Lorentz group, and the group variables $g_{ev}$ belong to $SL(2,\mbb C)$, the double cover of the proper orthochronous subgroup $SO^+(1,3)$. We still have $SU(2)$ spins $j_f$, with $d_{j_f}=2j_f+1$, and Bloch $SU(2)$ coherent states labeled by unit 3-vectors $\vec n_{ef}$.

The group integrals are performed with the Haar measure of $SL(2,\mbb C)$ but the redundant integrations (infinite volumes of $SL(2,\mbb C)$) have to be dropped in order to make the amplitude potentially finite \cite{Engle:2008ev}. This amounts to dropping one $SL(2,\mbb C)$ integral per each vertex.

The map $Y$ is constructed using the principal series of unitary representations $\mathcal H^{(j,\rho)}$ of $SL(2,\mbb C)$, which are infinite dimensional and labeled by a half-integer $j$ and a positive real number $\rho$. The spinfoam model imposes the constraint
\begin{align}\label{rhoeqgammak}
\rho=\gamma k
\end{align}
on the label $\rho$. This is the analogue of \eqref{ratio}. Notice that the constraint \eqref{rhoeqgammak} does not imply a quantization of the Barbero-Immirzi parameter as in the Euclidean case. This time it can be any nonzero real number; this is a comforting feature also from the Hamiltonian loop quantum gravity standpoint, since in the canonical theory $\gamma$ is not quantized. Moreover, we have the useful orthogonal decomposition
\begin{align}\label{decompo}
\mathcal H^{(j,\rho)}=\bigoplus_{k\geq j}\mathcal H^{(j,\rho)}_k,
\end{align}
where the subspace $\mathcal H^{(j,\rho)}_k$ carries an irreducible representation of spin $k$ of the $SU(2)$ subgroup of $SL(2,\mbb C)$ that leaves a reference timelike 4-vector invariant. Thus, similarly to the Euclidean signature, we need a fiducial basis in order to define concretely the $SU(2)$ sub-representations. In particular we need a time direction in $\mbb R^{1,3}$. The map $Y$ in \eqref{PefLOR} is then defined in terms of the decomposition \eqref{decompo} as the isometric embedding of the $SU(2)$ irreducible representation $j$ into the lowest weight subspace of the $SL(2,\mbb C)$ irreducible representation $\mathcal H^{(j,\gamma j)}$, namely the $SU(2)$-irreducible subspace with $k=j$. This completes the definition of the Lorentzian model.

In order to cast the model in a suitable exponential form, it is useful to realize the Hilbert space $\mathcal H^{(j,\rho)}$ in the common way as the space of homogeneous complex-valued functions on $\mbb C^2$,
\begin{align}
f(\lambda z)=\lambda^{-1+i\rho+j}\bar\lambda^{-1+i\rho-j}f(z),
\end{align}
with group transpose action
\begin{align}\label{transposeaction}
gf(\lambda z):=f(g^t z),
\end{align}
and $SL(2,\mbb C)$-invariant scalar product
\begin{align}
(f,g)=\int_{\mbb C\mbb P^1}dz\,\overline{f(z)}g(z).
\end{align}
The integration over the complex projective line $\mbb C\mbb P^1$ is well-defined thanks to the correct homogeneity of the measure
\begin{align}
dz=\frac{i}{2}(z^0dz^1-z^1dz^0)\wedge(\bar z^0d\bar z^1-\bar z^1d\bar z^0).
\end{align}
Thus $z=(z_1,z_2)\in\mbb C^2$ can be interpreted as homogeneous coordinates $[z_1:z_2]$ of $\mbb C\mbb P^1$.

Now we use the properties of the Bloch coherent states and of the map $Y$ in this realization, for which we refer the reader to the technology developed in the single-vertex analysis \cite{Barrett:2009mw}. Very briefly, a Bloch state $\ket{j, \vec n}\in\mathcal H^j$ is realized as the function 
\begin{align}
f^j_\xi(z)\propto\langle \bar z,\xi_{\vec n}\rangle^{2j},
\end{align}
where $\xi_{\vec n}$ is just the Bloch state in the fundamental $1/2$ representation, viewed as a $\mbb C^2$ vector, and the proportionality factor is determined by normalization. Under $Y$, the Bloch state is mapped to the state in $\mathcal H^{(j,\gamma j)}$ given by
\begin{equation}\label{Ymapped}
Y f^j_\xi(z)\propto\langle z,z\rangle^{-1-j-i\gamma j}\langle\bar z,\xi_{\vec n}\rangle^{2j}.
\end{equation}
In the following we will write $\xi$ instead of $\xi_{\vec n}$ for simplicity.

Thus using \eqref{Ymapped} into \eqref{PfLOR}, together with the group action \eqref{transposeaction} and, for convenience, a change of variable $z\rightarrow\bar z$ in the integral, the piece of the face amplitude associated to a single vertex can be written as
\begin{align}\label{PefKernel}
P_{vf}=\int dz\,\omega(z)\big(P_{vf}^0\big)^{i\gamma j_f}\big(P'_{vf}\big)^{j_f},
\end{align}
where $\omega(z)$ is slowly varying in the spin $j_f$ and independent of $\gamma$, and
\begin{align}
&P_{vf}^0=\frac{\|g^\dagger_{ve}z\|^2}{\|g^\dagger_{ve'}z\|^2},\\
&P'_{vf}=\frac{\langle \xi_{e'f},g^\dagger_{ve'}z\rangle^{2}\langle g^\dagger_{ve}z, \xi_{ef}\rangle^{2}}{\|g^\dagger_{ve'}z\|^2\,\|g^\dagger_{ve}z\|^2}.
\end{align}
Here $e=s(vf)$ and $e'=t(vf)$ are the two adjacent edges of $f$ sharing the vertex $v$. This allows us to define the action
\begin{align}\label{totalSlor}
S=S^0+\frac{1}{\gamma}S',
\end{align}
with
\begin{align}
&S^0=i\sum_{f}\sum_{v\in f}a_f\log P^0_{vf},\\
&S'=\sum_{f}\sum_{v\in f}a_f\log P'_{vf}.
\end{align}
Thus we rewrite the Lorentzian spinfoam amplitude \eqref{LorZ} in exponential form in terms of the action \eqref{totalSlor} as
\begin{align}
W=\sum_{\{a_f\}}\int \!dg_{ev}\!\int \!d\vec n_{ef}\! \int dz_{vf} \,\Omega\,e^{S},
\end{align}
with measure density given by 
\begin{align}
\Omega=\prod_{f}d_{j_f}\prod_{v\in f}\omega_{vf}.
\end{align}
Observe that we have one integration $dz_{vf}$ on $\mbb C\mbb P^1$ per each wedge $vf$ of the 2-complex. The action \eqref{totalSlor} and the variables used are a simple generalization to an arbitrary 2-complex of the single-vertex action studied in \cite{Barrett:2009mw}. Notice also that $S^0$ is purely imaginary whereas $S'$ is complex.

As we did for the Euclidean case, in order to find the dominant configurations for $\gamma\rightarrow 0$ we need to study the critical points of the complex action $S'$, because $S'$ is the piece of the action multiplied by the large parameter $1/\gamma$. The critical points are the stationary points satisfying a real part condition. Notice that in order to take the variations with respect to the areas we suppose that the sum over the spins can be approximated with an integral, with an error small in $\gamma$.

Imposing $\text{Re}\,S'=0$ and the vanishing of the variation of $S'$ with respect to $z_{vf}$ one finds the matching conditions
\begin{align}\label{LORcrit1}
g_{ve}\xi_{ef}&=e^{i\phi_{vf}}\frac{\|g^\dagger_{ve} z_{vf}\|^2}{\|g^\dagger_{ve'} z_{vf}\|^2} g_{ve'} \xi_{e'f},\\\label{LORcrit2}
(g^\dagger_{ve})^{-1}\xi_{ef}&=e^{i\phi_{vf}}\frac{\|g^\dagger_{ve'} z_{vf}\|^2}{\|g^\dagger_{ve} z_{vf}\|^2} (g^\dagger_{ve'})^{-1} \xi_{e'f},
\end{align}
similar to the Euclidean \eqref{critical}, where $e=s(vf)$ and $e'=t(vf)$ are the two adjacent edges in the boundary of the face $f$ and $v$ bounds both edges.

The requirement of stationarity with respect to $\vec n_{ef}$ gives an empty equation. Varying with respect to $g_{ev}$ and using the previous critical equations we get the closure condition
\begin{align}\label{LORclosure}
\left.\delta_{g_{ev}} S'\right|_{crit.}=0\Rightarrow\sum_{f\in e}\epsilon_{vef} a_{f}\vec n_{ef}=0
\end{align}
for each tetrahedron $e^*$. The variation of $S'$ with respect to the areas $a_f$ will be considered later. Let us first discuss the geometrical interpretation.

The solutions of the previous critical conditions \eqref{LORcrit1}, \eqref{LORcrit2}, \eqref{LORclosure} can be interpreted as 4-dimensional Lorentzian Regge triangulations, using the following construction. Given a critical configuration $(a_f, \vec n_{ef}, g_{ev})$ we define two types of null (lightlike) 4-vectors
\begin{align}
N_{vef}&=\frac{1}{2}g_{ve}(1,-\vec n_{ef}),\\
\tilde N_{vef}&=\frac{1}{2}g_{ve}(1,\vec n_{ef}).
\end{align}
Taking their wedge product, we define the spacelike simple area bivectors
\begin{align}\label{areabivLOR}
A_{vf}=2 a_f (N_{evf} \wedge \tilde N_{evf})^*
\end{align}
with norm $\|A_{vf}\|=a_f$, where we have used the fiducial Hodge operator. Notice that we have dropped the edge label $e$ of the area bivector because \eqref{LORcrit1} and \eqref{LORcrit2} imply the matching condition $A_{vef}=A_{ve'f}=:A_{vf}$. Thus for each vertex $v$ we have ten bivectors $A_{vf}$ constructed in this manner.

From simplicity, cross-simplicity, closure and the hypothesis of non-degeneracy we can use the Lorentzian version of the Barrett-Crane 4-simplex reconstruction theorem: there is a unique oriented geometric 4-simplex in $\mbb R^{1,3}$ with ten area bivectors equal to $A_{vf}$. The norm of the area bivectors is the Lorentzian triangle area, computed with the Minkowski metric $\eta_{\mu\nu}$. Notice that the reconstructed 4-simplex has a spacelike boundary because all $A_{vf}$ are spacelike, namely each tetrahedron in the boundary of the 4-simplex lives in a spacelike 3-plane.

Two neighboring geometric 4-simplices $v^*$ and $v'^*$ are described by two sets of ten bivectors $A_{vf}$ and $A_{v'f}$. The shape of the tetrahedron $e^*$, the dual to the edge connecting the two 4-simplices, is the same in the two 4-simplices, because the three-dimensional geometry of the tetrahedron is described by the common set of four normals $\epsilon_{vef}a_f \vec n_{ef}=-\epsilon_{v'ef}a_f \vec n_{ef}$. Therefore there exists a $O(1,3)$ Levi-Civita connection $U_{v'v}$ connecting the inertial frame at $v$ with the one at $v'$.

Thus the on-shell bivectors $A_{vf}$ defined in \eqref{areabivLOR} are the area bivectors of a Lorentzian triangulation, with Levi-Civita connection given by formula \eqref{LCwithSFgroup1} or \eqref{LCwithSFgroup2}, depending on the relative 4-simplex orientation.

Similarly to the Euclidean theory, we are interested in the geometrical interpretation of the quantities
\begin{align}
&\tilde\Theta_f:=\sum_{v\in f}\log \frac{\|g^\dagger_{ve'} z_{vf}\|^2}{\|g^\dagger_{ve} z_{vf}\|^2},\\
&\Theta_f^*:=\sum_{v\in f}\phi_{vf},
\end{align}
which are the angles appearing when we evaluate on a critical configuration the action terms $S^0$ and $S'$, respectively. These are also the angles in the expression of the spinfoam loop holonomy around the face:
\begin{align}\label{sfloopLor}
U(g_{vf})=e^{\tilde\Theta_f B_{vf}+\Theta_f^*B^*_{vf}},
\end{align}
where the rotation on the plane of $B_{vf}:=A_{vf}^*$ is a boost on the plane orthogonal to the triangle $f^*$ of the geometric 4-simplex $v^*$, and the rotation on the plane of $B^*_{vf}$ is a spatial rotation on the plane of the triangle. In \eqref{sfloopLor} the bivectors are viewed as Lorentz algebra elements using $\Lambda^2 \mbb R^{1,3}\simeq so(1,3)$.

The geometrical meaning of $\tilde\Theta_f$ is explained in terms of dihedral angles. The boundary of a 4-simplex is constituted of five spacelike tetrahedra, and it can be shown (see \cite{Barrett:2009mw} for a proof) that on the critical configurations the real ratio 
\begin{align}\label{Lordihedral}
\frac{\|g^\dagger_{ve'} z_{vf}\|^2}{\|g^\dagger_{ve} z_{vf}\|^2}=e^{\epsilon_v\Theta_{vf}}
\end{align}
provides the Lorentzian dihedral angle $\Theta_{vf}$ between the two tetrahedra $e^*$ and $e'^*$ of the reconstructed geometric 4-simplex $v^*$, where $e=s(vf)$ and $e'=t(vf)$. It is called dihedral `angle' but it is a dihedral boost parameter. It is defined up to a sign by
\begin{align}
\cosh\Theta_{vf}=|N_{s(vf)}\cdot N_{t(vf)}|,
\end{align}
where $N_e$ is the unit timelike outward normal to the tetrahedron $e^*$. The sign of the Lorentzian dihedral angle is defined as follows. If one normal is future pointing and the other past pointing (i.e. $vf$ is a thin wedge) the dihedral angle is defined as positive, otherwise (thick wedge) it is defined negative.\footnote{Notice that the positiveness of the dihedral angle $\Theta_{vf}$ has nothing to do with the orientation sign $\epsilon_v$ in front of it in \eqref{Lordihedral}.} By \eqref{Lordihedral}, $\tilde\Theta_f$ is thus the generalized deficit angle, which is orientation-dependent.

On the other hand $\Theta_f^*$ is a torsion angle, because it is the rotation on the plane of the triangle resulting from the parallel transport around the face. Because of its parity-invariance it is orientation-independent. Thus, as we did in the Euclidean analysis, we can compute it on a critical configuration with constant 4-simplex orientations without loosing generality. Given that in this case the loop spinfoam holonomy is equal to the loop Levi-Civita holonomy up to the spacetime inversion factor $\epsilon_f$, we must have
\begin{align}\label{ThetastarcritLor}
\Theta_f^*=
\begin{cases}
0&\epsilon_f=1\\
\pi&\epsilon_f=-1
\end{cases}
\end{align}
where the second case arises for an odd number of thick wedges of $f$, or equivalently when the loop Levi-Civita holonomy $U_{vf}$ is not orthochronous (time-unoriented face). The previous equation expresses the fact that the on-shell connection is torsion-free, as expected on-shell.

We can finally consider the last stationary phase equation arising by the variation of $S'$ with respect to the areas $a_f$. The result is identical to the Euclidean \eqref{areavariation}. Namely, for $\epsilon_f=1$ it gives an empty equation, whereas for $\epsilon_f=-1$ the resulting condition cannot be satisfied. Hence the geometries that are not time-oriented over all the triangulation are not critical and thus suppressed in the flipped limit. (We stress again that the suppression of the time-unoriented configurations was first noticed in \cite{Han:2013ina})

Now we can give the final expression of the on-shell action. On a critical configuration the term $S'$ of the full action vanishes, thanks to the torsionless constraint \eqref{ThetastarcritLor} and the time-orientedness. Therefore the on-shell total action is given by
\begin{align}\label{onshLOR}
S|_{crit.}=i S_R=i \sum_f a_f \tilde\Theta_{f},
\end{align}
which is the generalized Regge action for Lorentzian gravity, where $\tilde\Theta_f$ is the generalized Lorentzian deficit angle. It is worth observing again that the critical equations force the areas $a_f$ to be Regge-like. The non-Regge-like areas are suppressed in the flipped limit.

From this point the study of the flipped and classical limits proceeds as in the Euclidean case, so we refer to the previous section for more details.
\section{Boundary states} The predictions of the spinfoam theory are the transition amplitudes\footnote{Other terms for the transition amplitudes are boundary amplitudes and extraction amplitudes \cite{Ashtekar:2010gz}. The latter is motivated by the fact that the spinfoam dynamics is supposed to `extract' the physical states, namely the quantum states that solve the Hamiltonian constraint, from the kinematical state space (see \cite{Noui:2004iy} for a proof in three-dimensional gravity). In other words the extraction amplitude `projects' the boundary state into a physical state.} for the states of 3-geometry defined on the boundary of the 2-complex \cite{Oeckl:2003vu}. So far the boundary of the 2-complex was always understood. In this section we take the boundary into account more explicitly and comment on its role in the classical limit.

The boundary of a 2-complex $\sigma$ is a 1-complex, that is a graph $\Gamma$ with nodes $p$ connected by links $l$. The links bound the external faces of the 2-complex and the nodes bound the external edges. The boundary graph inherits the orientation from the external faces. Though many arguments in this paper are for general 2-complexes, we work mainly with 2-complexes defined on a simplicial triangulation; thus the nodes of the boundary graph are 4-valent and dual to tetrahedra, and the links are dual to triangles.

The transition amplitudes can be expressed in different bases. For the models considered in this paper (Euclidean EPRL-FK with $0<\gamma<1$ and Lorentzian EPRL for arbitrary $\gamma$) the boundary Hilbert space of 3-geometry is the one of Hamiltonian loop quantum gravity,
\begin{align}\label{statespace}
\mathcal H_\Gamma=\mathcal L^2(SU(2)^L/SU(2)^N),
\end{align}
where $L$ is the number of links of $\Gamma$, $N$ the number of nodes, and the quotient means that the space $\mathcal H_\Gamma$ is obtained from $\mathcal L^2(SU(2)^L)$ by projecting at each note on the gauge-invariant subspace. So $\mathcal H_\Gamma$ is the Hilbert space of gauge-invariant square-integrable functions of $L$ copies of $SU(2)$ (with respect to the Haar measure).

If $\Gamma$ has two components $\Gamma=\Gamma_{in}\cup\Gamma_{out}$ we may interpret the amplitude as a standard transition amplitude from an initial state in $\mathcal H_{\Gamma_{in}}$ to a final state in $\mathcal H_{\Gamma_{out}}$.

In the state space \eqref{statespace} there exists an overcomplete basis of spin-networks labeled by spins $j_l$ (on the links) and $SU(2)$ elements $n_{pl}$ (on the nodes, one per each link $l$ bounded by $p$), introduced in quantum gravity by Livine and Speziale \cite{Livine:2007vk}. Remember that a $SU(2)$ element $n$ can be thought as a unit vector $\vec n = n \hat z$ up to a $U(1)$ phase ambiguity, where $\hat z=(0,0,1)$ is a reference direction in $\mbb R^3$. A choice of this arbitrary phase is understood in the following.

In this spin-network basis, the transition amplitude is a function of the previous labels, namely
\begin{align}\label{boundaryfunctional}
W(j_l,\vec n_{pl})=\langle W|j_l,\vec n_{pl}\rangle.
\end{align}
Notice that these are the same labels we used in the previous sections for the dynamical variables in the bulk of the triangulation. In a sense we can use the spin-network basis on the boundary (as input boundary data) \emph{and} in the bulk (as intermediate states being summed over), very much like in the Feynman diagrams.

Taking explicitly into account the boundary data, the spinfoam amplitude \eqref{amplitudejn} now reads
\begin{align}\label{amplitudejnSS}
W(j_l,\vec n_{pl})=\sum_{\{j_f\}}\int\!dg_{ev}\!\int d\vec n_{ef}\prod_l P_l\prod_f d_{j_f}P_f,
\end{align}
where the face amplitude is split into a boundary part relative to the external faces (or equivalently to the links $l$ on the boundary) and a bulk part relative to the internal faces $f$. Accordingly, the two products in \eqref{amplitudejnSS} are over the external and internal faces, respectively.

Let us explain better the last expression. The external face amplitude $P_l$ has the same form of the internal $P_f$ (see \eqref{Pf}) except that for the external edges of the face we have only `half' edge-face operator. More precisely, we have
\begin{align}
P_{ef}=\bra{j_f,\vec n_{ef}}Y^\dagger g_{e,s(ef)}
\end{align}
or
\begin{align}
P_{ef}=g_{t(ef),e}Y\ket{j_f,\vec n_{ef}}
\end{align}
depending on the orientation of the external edge induced by the orientation of the external face. The sum in \eqref{amplitudejnSS} is over the internal spins. The integrals are over the internal unit vectors and the group variables ($Spin(4)$ in the Euclidean and $SL(2,\mbb C)$ in the Lorentzian). The external spins label the external faces, or equivalently the links $l$ of the boundary graph $\Gamma$; hence for them we can use the notation $j_l$. The external unit vectors label the external edges, or equivalently the nodes $p$ of $\Gamma$, and we can use the notation $\vec n_{pl}$. In this way the transition amplitude $W(j_l,\vec n_{pl})$ is a function of the boundary labels.

All the previous analysis of the flipped and classical limits is to be understood in the context of the transition amplitudes \eqref{boundaryfunctional} with boundary states. Before we did not write explicitly the contribution of the boundary in order not to overcomplicate the formulas unnecessarily. For example, the exponential form of the Lorentzian amplitude for a 2-complex with boundary reads
\begin{align}\label{amplitudejnSSLOR}
W(j_l,\vec n_{pl})=\sum_{\{j_f\}}\!\int dg_{ev}\!\int\! d\vec n_{ef}\!\int\! dz_{vf}\,\Omega \,e^{\sum S_l+S_f}.
\end{align}

The intermediate states of quantum 4-geometry interpolate the quantum 3-geometry represented by the boundary data. The boundary state fixes the scale of the geometry we want to study. When this scale is large with respect to the Planck scale we select the classical regime of the spinfoam transition amplitude. Our main working hypothesis, necessary for the setup of the stationary phase method, is that the intermediate states with areas of the same order of the boundary areas are the ones contributing the most to the amplitude in the regime studied.

Though checking the correctness of this hypothesis could be difficult in general, one could be satisfied with a check \emph{a posteriori} of the validity of the approximation, after the formal asymptotic expansion. We were not able to provide this check in the paper. Nevertheless, we point out that there is substantial evidence in the literature that the hypothesis is correct (see in particular \cite{Bianchi:2008ae}).

In order to state the regime we have studied in the previous sections in terms of the transition amplitudes with boundary states let us write with an abuse of notation the amplitude as a function of the boundary areas instead of the spins as $W(a_l,\vec n_{pl})$. The regime we are interested in is large boundary areas $a_l$, much larger than the Planck area (classical limit) but much smaller than $a_l/\gamma$ (flipped limit of small $\gamma$).

As we have seen previously, in this regime the dominant contribution of the spinfoam amplitude is from configurations that solve the Einstein equations \eqref{Einsteintetrad}, or better their simplicial version provided by the generalized Regge equations \eqref{Ricciflat}. These classical configurations interpolate the boundary data and thus the role of the boundary state is to select the bulk four-dimensional geometries which are compatible with the boundary three-dimensional geometry specified (see \cite{Sorkin:1975ah,Barrett:1994ks} on the initial value problem). As we know from classical Regge gravity, the nature and number of the solutions depends on the structure of the triangulation and its boundary data. The same is true in spinfoams.

Thus we have motivated the following behavior of the spinfoam transition amplitude in the combined classical regime and small Barbero-Immirzi parameter. If there is at most one solution of the generalized Regge equations the asymptotic amplitude has the form
\begin{align}\label{result}
W(a_l,\vec n_{pl})\sim \sum_\epsilon A_{\epsilon}\,e^{\frac{i}{\hbar}S^\epsilon_R(a_l,\vec n_{pl})},
\end{align}
where now $S^\epsilon_R$ is the Hamilton function, that is the generalized Regge action evaluated on the classical solution determined by the boundary data $(a_l,\vec n_{pl})$.\footnote{The prefactors $A_\epsilon$ are non-oscillatory and slowly varying.} In the case there are many solutions to the generalized Regge equations (for fixed joint orientations $\epsilon$), their respective contributions must be added to \eqref{result}.

Finally, we observe that the boundary data $(a_l,\vec n_{pl})$ may fail to be Regge-like at a three-dimensional level. Namely there may not exist a three-dimensional Regge triangulation with the prescribed areas and unit vectors, the latter interpreted as the three-dimensional unit normals of the tetrahedra in the boundary. From the geometrical interpretation of the critical configurations it is clear that in this case the transition amplitude would be exponentially suppressed. On the other hand, a consistent Regge-like set of boundary data can be mapped into the equivalent set of lengths of the boundary triangulation. This set of lengths is a Dirichlet boundary condition for the generalized Regge equations of motion to be used to determine the classical solution in the interior and evaluate the Hamilton function in \eqref{result}.

The last formula \eqref{result} is a concrete realization of the equation \eqref{whatweexpect} in the introduction.
\section*{Conclusions and outlook}
It this paper we have discussed a proposal for the classical limit of Euclidean EPRL-FK and Lorentzian EPRL spinfoams truncated to an arbitrary, finite triangulation. We find (equation \eqref{result}) that the transition amplitudes yield the exponential of the Hamilton function of general relativity, up to corrections in $l_P^2/a$ and $\gamma$-corrections in $\gamma l_P^2/a$, in the regime
\begin{align}\label{compactregimebis}
l^2_P\ll a\ll a/\gamma,
\end{align}
namely in the simultaneous classical limit of large areas and flipped limit of small Barbero-Immirzi parameter, taken in the appropriate order.

The result can be explained by observing that the Barbero-Immirzi parameter controls the strength of the geometricity constraints at the quantum level, in particular the torsionless constraint, and for $\gamma\rightarrow 0$ the first-order quantum theory reduces to an effective second-order theory similar to quantum Regge gravity. In turn, the classical solutions of Regge gravity found in the classical limit of large geometries satisfy the simplicial version of the Einstein equations.

However, it is quite surprising that one does not get the correct classical limit by just varying the full spinfoam action $S=S^0+S'/\gamma$, which corresponds to looking at the large area limit with $\gamma$ fixed. Indeed in such regime one would get only the flat spacetime solutions \cite{Bonzom:2009hw, Hellmann:2013gva, Perini:2012nd}, and this is not compatible with the Einstein equations. This is suspect and seems to hint to a problem with the EPRL and FK models because, on the contrary, the variation of the full Holst action \eqref{Sintro} with respect to $e$ and $\omega$ does yield the Einstein equations. It means, basically, that the fluctuations of the spinfoam variables are more general than the fluctuations of $e$ and $\omega$.

On the other hand the fact that for small Barbero-Immirzi parameter the Einstein equations seem to be correctly reproduced opens the way to some other interpretation of the result, though still very speculative. For example we have suggested that the mechanism behind the flipped regime could be related to the coarse-graining of the triangulation. Maybe the effective spinfoam amplitude for a smaller triangulation (with less 4-simplices, not smaller in the overall physical size!) renormalizes the Barbero-Immirzi parameter to smaller values. This interpretation of the flipped limit is not yet clear to the authors and deserves further study.

Our analysis sheds new light on the previous calculations of the graviton propagator and other correlation functions \cite{Bianchi:2009ri,Bianchi:2011hp,Rovelli:2011kf} where similar results hold at a single 4-simplex level. We believe that these results on the n-point functions are now much more clear.

Interestingly, a limit resembling the one studied here was considered by Bojowald \cite{Bojowald:2001ep} in the context of loop quantum cosmology. Quoting the abstract: ``\emph{standard quantum cosmology is shown to be the simultaneous limit $\gamma\rightarrow 0$, $j\rightarrow\infty$ of loop quantum cosmology}''. It would be nice to study this analogy more in detail.

We conclude with some comments on what we believe is still missing in the present analysis.

First, we stress again that the main result of the present paper is conditional on few hypothesis. Our main working hypothesis is that the configurations with internal areas of the same order of the boundary areas are the main contribution to the path integral in the regime studied. This hypothesis was necessary in order to justify the use of the stationary phase method for the internal action.

We do not know at this stage whether the terms with mixed orientations $\epsilon$ in the final formula \eqref{result} are non-vanishing. To give an answer, one should study the solutions of the generalized Regge equations. Though we expect these solutions to contribute on the same footing as the solutions of the proper Regge equations, the answer is still not completely clear and to the authors' knowledge little is known on this topic in the literature, at least in the simplicial setting. Therefore we leave this interesting point for a future work.

Nevertheless, there is good evidence that the transition amplitudes for semiclassical states\footnote{A semiclassical boundary state is peaked on both the intrinsic and extrinsic geometry of the boundary, similarly to a particle semiclassical state, which is peaked on both position and momentum, with small relative dispersions.} are able to select dynamically the globally oriented configurations (see \cite{Bianchi:2008ae} for a comprehensive discussion in the context of Regge calculus and \cite{Bianchi:2010mw} for a single-vertex discussion in spinfoams). In the Lorentzian setting, the definition of causal amplitudes makes use of modified spinfoam amplitudes which are orientation-dependent \cite{Livine:2002rh,Oriti:2004yu,Oriti:2006wq}.

Another missing piece in our analysis is the study of the degenerate sector of the amplitude in the flipped limit. We have disregarded the critical configurations representing degenerate geometries and their possible contribution to the final asymptotic formula. A careful study of these terms is desirable.

Finally, we have disregarded the potential `infrared' divergencies associated to the bubbles of the foam. The bubbles are present for certain types of 2-complexes and are similar to the loops of perturbative QFT (see \cite{Perini:2008pd,Krajewski:2010yq,Geloun:2010vj,Rivasseau:2011xg}). A physical regulator such as a nonvanishing cosmological constant \cite{Han:2010pz, Fairbairn:2011aa} could be required to make the results presented here more rigorous and general.
\section*{Acknowlegements}
We are grateful to Martin Bojowald, Laurent Freidel, Frank Hellmann and Carlo Rovelli for useful comments. We thank the organizers of the Loops 11 Conference held in Madrid, where some of these ideas were sharpened thanks to enlightening discussions. This work was supported in part by the NSF grant PHY0854743, The George A. and Margaret M.~Downsbrough Endowment and the Eberly research funds of Penn State. E.M. gratefully acknowledges support from ``Fondazione Angelo della Riccia''.

The last revision of this paper was completed in the sunny village of Valbonne, in the South of France, after a long work of polishing. We have corrected many typos and imprecisions, and largely improved the presentation. Though the core ideas and results are unchanged, some more technical details were missing in the previous versions. In particular, in order to make the paper self-contained, we have added the discussion of the time-orientedness of geometry found recently by Muxin Han, and we thank him for a useful exchange on this point.
\bibliography{biblioreggefromsf}

\begin{thebibliography}{66}%
\makeatletter
\providecommand \@ifxundefined [1]{%
 \@ifx{#1\undefined}
}%
\providecommand \@ifnum [1]{%
 \ifnum #1\expandafter \@firstoftwo
 \else \expandafter \@secondoftwo
 \fi
}%
\providecommand \@ifx [1]{%
 \ifx #1\expandafter \@firstoftwo
 \else \expandafter \@secondoftwo
 \fi
}%
\providecommand \natexlab [1]{#1}%
\providecommand \enquote  [1]{``#1''}%
\providecommand \bibnamefont  [1]{#1}%
\providecommand \bibfnamefont [1]{#1}%
\providecommand \citenamefont [1]{#1}%
\providecommand \href@noop [0]{\@secondoftwo}%
\providecommand \href [0]{\begingroup \@sanitize@url \@href}%
\providecommand \@href[1]{\@@startlink{#1}\@@href}%
\providecommand \@@href[1]{\endgroup#1\@@endlink}%
\providecommand \@sanitize@url [0]{\catcode `\\12\catcode `\$12\catcode
  `\&12\catcode `\#12\catcode `\^12\catcode `\_12\catcode `\%12\relax}%
\providecommand \@@startlink[1]{}%
\providecommand \@@endlink[0]{}%
\providecommand \url  [0]{\begingroup\@sanitize@url \@url }%
\providecommand \@url [1]{\endgroup\@href {#1}{\urlprefix }}%
\providecommand \urlprefix  [0]{URL }%
\providecommand \Eprint [0]{\href }%
\providecommand \doibase [0]{http://dx.doi.org/}%
\providecommand \selectlanguage [0]{\@gobble}%
\providecommand \bibinfo  [0]{\@secondoftwo}%
\providecommand \bibfield  [0]{\@secondoftwo}%
\providecommand \translation [1]{[#1]}%
\providecommand \BibitemOpen [0]{}%
\providecommand \bibitemStop [0]{}%
\providecommand \bibitemNoStop [0]{.\EOS\space}%
\providecommand \EOS [0]{\spacefactor3000\relax}%
\providecommand \BibitemShut  [1]{\csname bibitem#1\endcsname}%
\let\auto@bib@innerbib\@empty
\bibitem [{\citenamefont {Baez}(1998)}]{Baez:1997zt}%
  \BibitemOpen
  \bibfield  {author} {\bibinfo {author} {\bibfnamefont {John~C.}\ \bibnamefont
  {Baez}},\ }\bibfield  {title} {\enquote {\bibinfo {title} {{Spin foam
  models}},}\ }\href {\doibase 10.1088/0264-9381/15/7/004} {\bibfield
  {journal} {\bibinfo  {journal} {Class. Quant. Grav.}\ }\textbf {\bibinfo
  {volume} {15}},\ \bibinfo {pages} {1827--1858} (\bibinfo {year} {1998})},\
  \Eprint {http://arxiv.org/abs/gr-qc/9709052} {arXiv:gr-qc/9709052}
  \BibitemShut {NoStop}%
\bibitem [{\citenamefont {Reisenberger}\ and\ \citenamefont
  {Rovelli}(2000)}]{Reisenberger:2000fy}%
  \BibitemOpen
  \bibfield  {author} {\bibinfo {author} {\bibfnamefont {Michael}\ \bibnamefont
  {Reisenberger}}\ and\ \bibinfo {author} {\bibfnamefont {Carlo}\ \bibnamefont
  {Rovelli}},\ }\bibfield  {title} {\enquote {\bibinfo {title} {{Spin foams as
  Feynman diagrams}},}\ }\href@noop {} {\  (\bibinfo {year} {2000})},\ \Eprint
  {http://arxiv.org/abs/gr-qc/0002083} {arXiv:gr-qc/0002083} \BibitemShut
  {NoStop}%
\bibitem [{\citenamefont {Perez}(2003)}]{Perez:2003vx}%
  \BibitemOpen
  \bibfield  {author} {\bibinfo {author} {\bibfnamefont {Alejandro}\
  \bibnamefont {Perez}},\ }\bibfield  {title} {\enquote {\bibinfo {title}
  {{Spin foam models for quantum gravity}},}\ }\href@noop {} {\bibfield
  {journal} {\bibinfo  {journal} {Class. Quant. Grav.}\ }\textbf {\bibinfo
  {volume} {20}},\ \bibinfo {pages} {R43} (\bibinfo {year} {2003})},\ \Eprint
  {http://arxiv.org/abs/gr-qc/0301113} {arXiv:gr-qc/0301113} \BibitemShut
  {NoStop}%
\bibitem [{\citenamefont {Misner}(1957)}]{Misner:1957wq}%
  \BibitemOpen
  \bibfield  {author} {\bibinfo {author} {\bibfnamefont {Charles~W.}\
  \bibnamefont {Misner}},\ }\bibfield  {title} {\enquote {\bibinfo {title}
  {{Feynman quantization of general relativity}},}\ }\href {\doibase
  10.1103/RevModPhys.29.497} {\bibfield  {journal} {\bibinfo  {journal} {Rev.
  Mod. Phys.}\ }\textbf {\bibinfo {volume} {29}},\ \bibinfo {pages} {497--509}
  (\bibinfo {year} {1957})}\BibitemShut {NoStop}%
\bibitem [{\citenamefont {Hawking}(1978)}]{Hawking:1979zw}%
  \BibitemOpen
  \bibfield  {author} {\bibinfo {author} {\bibfnamefont {S.~W.}\ \bibnamefont
  {Hawking}},\ }\bibfield  {title} {\enquote {\bibinfo {title} {{Space-Time
  Foam}},}\ }\href {\doibase 10.1016/0550-3213(78)90375-9} {\bibfield
  {journal} {\bibinfo  {journal} {Nucl. Phys.}\ }\textbf {\bibinfo {volume}
  {B144}},\ \bibinfo {pages} {349--362} (\bibinfo {year} {1978})}\BibitemShut
  {NoStop}%
\bibitem [{\citenamefont {Rovelli}\ and\ \citenamefont
  {Smerlak}(2012)}]{Rovelli:2010qx}%
  \BibitemOpen
  \bibfield  {author} {\bibinfo {author} {\bibfnamefont {Carlo}\ \bibnamefont
  {Rovelli}}\ and\ \bibinfo {author} {\bibfnamefont {Matteo}\ \bibnamefont
  {Smerlak}},\ }\bibfield  {title} {\enquote {\bibinfo {title} {{In quantum
  gravity, summing is refining}},}\ }\href {\doibase
  10.1088/0264-9381/29/5/055004} {\bibfield  {journal} {\bibinfo  {journal}
  {Class.Quant.Grav.}\ }\textbf {\bibinfo {volume} {29}},\ \bibinfo {pages}
  {055004} (\bibinfo {year} {2012})},\ \Eprint {http://arxiv.org/abs/1010.5437}
  {arXiv:1010.5437 [gr-qc]} \BibitemShut {NoStop}%
\bibitem [{\citenamefont {Engle}\ \emph
  {et~al.}(2008{\natexlab{a}})\citenamefont {Engle}, \citenamefont {Livine},
  \citenamefont {Pereira},\ and\ \citenamefont {Rovelli}}]{Engle:2007wy}%
  \BibitemOpen
  \bibfield  {author} {\bibinfo {author} {\bibfnamefont {Jonathan}\
  \bibnamefont {Engle}}, \bibinfo {author} {\bibfnamefont {Etera}\ \bibnamefont
  {Livine}}, \bibinfo {author} {\bibfnamefont {Roberto}\ \bibnamefont
  {Pereira}}, \ and\ \bibinfo {author} {\bibfnamefont {Carlo}\ \bibnamefont
  {Rovelli}},\ }\bibfield  {title} {\enquote {\bibinfo {title} {{LQG vertex
  with finite Immirzi parameter}},}\ }\href {\doibase
  10.1016/j.nuclphysb.2008.02.018} {\bibfield  {journal} {\bibinfo  {journal}
  {Nucl. Phys.}\ }\textbf {\bibinfo {volume} {B799}},\ \bibinfo {pages}
  {136--149} (\bibinfo {year} {2008}{\natexlab{a}})},\ \Eprint
  {http://arxiv.org/abs/0711.0146} {arXiv:0711.0146 [gr-qc]} \BibitemShut
  {NoStop}%
\bibitem [{\citenamefont {Livine}\ and\ \citenamefont
  {Speziale}(2008)}]{Livine:2007ya}%
  \BibitemOpen
  \bibfield  {author} {\bibinfo {author} {\bibfnamefont {Etera~R.}\
  \bibnamefont {Livine}}\ and\ \bibinfo {author} {\bibfnamefont {Simone}\
  \bibnamefont {Speziale}},\ }\bibfield  {title} {\enquote {\bibinfo {title}
  {{Consistently Solving the Simplicity Constraints for Spinfoam Quantum
  Gravity}},}\ }\href {\doibase 10.1209/0295-5075/81/50004} {\bibfield
  {journal} {\bibinfo  {journal} {Europhys. Lett.}\ }\textbf {\bibinfo {volume}
  {81}},\ \bibinfo {pages} {50004} (\bibinfo {year} {2008})},\ \Eprint
  {http://arxiv.org/abs/0708.1915} {arXiv:0708.1915 [gr-qc]} \BibitemShut
  {NoStop}%
\bibitem [{\citenamefont {Freidel}\ and\ \citenamefont
  {Krasnov}(2008)}]{Freidel:2007py}%
  \BibitemOpen
  \bibfield  {author} {\bibinfo {author} {\bibfnamefont {Laurent}\ \bibnamefont
  {Freidel}}\ and\ \bibinfo {author} {\bibfnamefont {Kirill}\ \bibnamefont
  {Krasnov}},\ }\bibfield  {title} {\enquote {\bibinfo {title} {{A New Spin
  Foam Model for 4d Gravity}},}\ }\href {\doibase
  10.1088/0264-9381/25/12/125018} {\bibfield  {journal} {\bibinfo  {journal}
  {Class. Quant. Grav.}\ }\textbf {\bibinfo {volume} {25}},\ \bibinfo {pages}
  {125018} (\bibinfo {year} {2008})},\ \Eprint {http://arxiv.org/abs/0708.1595}
  {arXiv:0708.1595 [gr-qc]} \BibitemShut {NoStop}%
\bibitem [{\citenamefont {Rovelli}(2011)}]{Rovelli:2010bf}%
  \BibitemOpen
  \bibfield  {author} {\bibinfo {author} {\bibfnamefont {Carlo}\ \bibnamefont
  {Rovelli}},\ }\bibfield  {title} {\enquote {\bibinfo {title} {{Loop quantum
  gravity: the first twenty five years}},}\ }\href {\doibase
  10.1088/0264-9381/28/15/153002} {\bibfield  {journal} {\bibinfo  {journal}
  {Class.Quant.Grav.}\ }\textbf {\bibinfo {volume} {28}},\ \bibinfo {pages}
  {153002} (\bibinfo {year} {2011})},\ \Eprint {http://arxiv.org/abs/1012.4707}
  {arXiv:1012.4707 [gr-qc]} \BibitemShut {NoStop}%
\bibitem [{\citenamefont {Rovelli}\ and\ \citenamefont
  {Speziale}(2011)}]{Rovelli:2010ed}%
  \BibitemOpen
  \bibfield  {author} {\bibinfo {author} {\bibfnamefont {Carlo}\ \bibnamefont
  {Rovelli}}\ and\ \bibinfo {author} {\bibfnamefont {Simone}\ \bibnamefont
  {Speziale}},\ }\bibfield  {title} {\enquote {\bibinfo {title} {{Lorentz
  covariance of loop quantum gravity}},}\ }\href {\doibase
  10.1103/PhysRevD.83.104029} {\bibfield  {journal} {\bibinfo  {journal}
  {Phys.Rev.}\ }\textbf {\bibinfo {volume} {D83}},\ \bibinfo {pages} {104029}
  (\bibinfo {year} {2011})},\ \Eprint {http://arxiv.org/abs/1012.1739}
  {arXiv:1012.1739 [gr-qc]} \BibitemShut {NoStop}%
\bibitem [{\citenamefont {Bianchi}\ \emph {et~al.}(2013)\citenamefont
  {Bianchi}, \citenamefont {Han}, \citenamefont {Rovelli}, \citenamefont
  {Wieland}, \citenamefont {Magliaro} \emph {et~al.}}]{Bianchi:2010bn}%
  \BibitemOpen
  \bibfield  {author} {\bibinfo {author} {\bibfnamefont {Eugenio}\ \bibnamefont
  {Bianchi}}, \bibinfo {author} {\bibfnamefont {Muxin}\ \bibnamefont {Han}},
  \bibinfo {author} {\bibfnamefont {Carlo}\ \bibnamefont {Rovelli}}, \bibinfo
  {author} {\bibfnamefont {Wolfgang}\ \bibnamefont {Wieland}}, \bibinfo
  {author} {\bibfnamefont {Elena}\ \bibnamefont {Magliaro}},  \emph {et~al.},\
  }\bibfield  {title} {\enquote {\bibinfo {title} {{Spinfoam fermions}},}\
  }\href {\doibase 10.1088/0264-9381/30/23/235023} {\bibfield  {journal}
  {\bibinfo  {journal} {Class.Quant.Grav.}\ }\textbf {\bibinfo {volume} {30}},\
  \bibinfo {pages} {235023} (\bibinfo {year} {2013})},\ \Eprint
  {http://arxiv.org/abs/1012.4719} {arXiv:1012.4719 [gr-qc]} \BibitemShut
  {NoStop}%
\bibitem [{\citenamefont {Han}\ and\ \citenamefont
  {Rovelli}(2013)}]{Han:2011as}%
  \BibitemOpen
  \bibfield  {author} {\bibinfo {author} {\bibfnamefont {Muxin}\ \bibnamefont
  {Han}}\ and\ \bibinfo {author} {\bibfnamefont {Carlo}\ \bibnamefont
  {Rovelli}},\ }\bibfield  {title} {\enquote {\bibinfo {title} {{Spin-foam
  Fermions: PCT Symmetry, Dirac Determinant, and Correlation Functions}},}\
  }\href {\doibase 10.1088/0264-9381/30/7/075007} {\bibfield  {journal}
  {\bibinfo  {journal} {Class.Quant.Grav.}\ }\textbf {\bibinfo {volume} {30}},\
  \bibinfo {pages} {075007} (\bibinfo {year} {2013})},\ \Eprint
  {http://arxiv.org/abs/1101.3264} {arXiv:1101.3264 [gr-qc]} \BibitemShut
  {NoStop}%
\bibitem [{\citenamefont {Rovelli}\ and\ \citenamefont
  {Smolin}(1995)}]{Rovelli:1994ge}%
  \BibitemOpen
  \bibfield  {author} {\bibinfo {author} {\bibfnamefont {Carlo}\ \bibnamefont
  {Rovelli}}\ and\ \bibinfo {author} {\bibfnamefont {Lee}\ \bibnamefont
  {Smolin}},\ }\bibfield  {title} {\enquote {\bibinfo {title} {{Discreteness of
  area and volume in quantum gravity}},}\ }\href {\doibase
  10.1016/0550-3213(95)00150-Q} {\bibfield  {journal} {\bibinfo  {journal}
  {Nucl.Phys.}\ }\textbf {\bibinfo {volume} {B442}},\ \bibinfo {pages}
  {593--622} (\bibinfo {year} {1995})},\ \Eprint
  {http://arxiv.org/abs/gr-qc/9411005} {arXiv:gr-qc/9411005 [gr-qc]}
  \BibitemShut {NoStop}%
\bibitem [{\citenamefont {Ding}\ and\ \citenamefont
  {Rovelli}(2010)}]{Ding:2010ye}%
  \BibitemOpen
  \bibfield  {author} {\bibinfo {author} {\bibfnamefont {You}\ \bibnamefont
  {Ding}}\ and\ \bibinfo {author} {\bibfnamefont {Carlo}\ \bibnamefont
  {Rovelli}},\ }\bibfield  {title} {\enquote {\bibinfo {title} {{Physical
  boundary Hilbert space and volume operator in the Lorentzian new spin-foam
  theory}},}\ }\href {\doibase 10.1088/0264-9381/27/20/205003} {\bibfield
  {journal} {\bibinfo  {journal} {Class. Quant. Grav.}\ }\textbf {\bibinfo
  {volume} {27}},\ \bibinfo {pages} {205003} (\bibinfo {year} {2010})},\
  \Eprint {http://arxiv.org/abs/1006.1294} {arXiv:1006.1294 [gr-qc]}
  \BibitemShut {NoStop}%
\bibitem [{\citenamefont {Bianchi}\ \emph {et~al.}(2006)\citenamefont
  {Bianchi}, \citenamefont {Modesto}, \citenamefont {Rovelli},\ and\
  \citenamefont {Speziale}}]{Bianchi:2006uf}%
  \BibitemOpen
  \bibfield  {author} {\bibinfo {author} {\bibfnamefont {Eugenio}\ \bibnamefont
  {Bianchi}}, \bibinfo {author} {\bibfnamefont {Leonardo}\ \bibnamefont
  {Modesto}}, \bibinfo {author} {\bibfnamefont {Carlo}\ \bibnamefont
  {Rovelli}}, \ and\ \bibinfo {author} {\bibfnamefont {Simone}\ \bibnamefont
  {Speziale}},\ }\bibfield  {title} {\enquote {\bibinfo {title} {{Graviton
  propagator in loop quantum gravity}},}\ }\href {\doibase
  10.1088/0264-9381/23/23/024} {\bibfield  {journal} {\bibinfo  {journal}
  {Class. Quant. Grav.}\ }\textbf {\bibinfo {volume} {23}},\ \bibinfo {pages}
  {6989--7028} (\bibinfo {year} {2006})},\ \Eprint
  {http://arxiv.org/abs/gr-qc/0604044} {arXiv:gr-qc/0604044} \BibitemShut
  {NoStop}%
\bibitem [{\citenamefont {Alesci}\ and\ \citenamefont
  {Rovelli}(2007)}]{Alesci:2007tx}%
  \BibitemOpen
  \bibfield  {author} {\bibinfo {author} {\bibfnamefont {Emanuele}\
  \bibnamefont {Alesci}}\ and\ \bibinfo {author} {\bibfnamefont {Carlo}\
  \bibnamefont {Rovelli}},\ }\bibfield  {title} {\enquote {\bibinfo {title}
  {{The complete LQG propagator: I. Difficulties with the Barrett-Crane
  vertex}},}\ }\href {\doibase 10.1103/PhysRevD.76.104012} {\bibfield
  {journal} {\bibinfo  {journal} {Phys. Rev.}\ }\textbf {\bibinfo {volume}
  {D76}},\ \bibinfo {pages} {104012} (\bibinfo {year} {2007})},\ \Eprint
  {http://arxiv.org/abs/0708.0883} {arXiv:0708.0883 [gr-qc]} \BibitemShut
  {NoStop}%
\bibitem [{\citenamefont {Bianchi}\ \emph {et~al.}(2009)\citenamefont
  {Bianchi}, \citenamefont {Magliaro},\ and\ \citenamefont
  {Perini}}]{Bianchi:2009ri}%
  \BibitemOpen
  \bibfield  {author} {\bibinfo {author} {\bibfnamefont {Eugenio}\ \bibnamefont
  {Bianchi}}, \bibinfo {author} {\bibfnamefont {Elena}\ \bibnamefont
  {Magliaro}}, \ and\ \bibinfo {author} {\bibfnamefont {Claudio}\ \bibnamefont
  {Perini}},\ }\bibfield  {title} {\enquote {\bibinfo {title} {{LQG propagator
  from the new spin foams}},}\ }\href {\doibase
  10.1016/j.nuclphysb.2009.07.016} {\bibfield  {journal} {\bibinfo  {journal}
  {Nucl. Phys.}\ }\textbf {\bibinfo {volume} {B822}},\ \bibinfo {pages}
  {245--269} (\bibinfo {year} {2009})},\ \Eprint
  {http://arxiv.org/abs/0905.4082} {arXiv:0905.4082 [gr-qc]} \BibitemShut
  {NoStop}%
\bibitem [{\citenamefont {Bianchi}\ \emph
  {et~al.}(2010{\natexlab{a}})\citenamefont {Bianchi}, \citenamefont
  {Rovelli},\ and\ \citenamefont {Vidotto}}]{Bianchi:2010zs}%
  \BibitemOpen
  \bibfield  {author} {\bibinfo {author} {\bibfnamefont {Eugenio}\ \bibnamefont
  {Bianchi}}, \bibinfo {author} {\bibfnamefont {Carlo}\ \bibnamefont
  {Rovelli}}, \ and\ \bibinfo {author} {\bibfnamefont {Francesca}\ \bibnamefont
  {Vidotto}},\ }\bibfield  {title} {\enquote {\bibinfo {title} {{Towards
  Spinfoam Cosmology}},}\ }\href {\doibase 10.1103/PhysRevD.82.084035}
  {\bibfield  {journal} {\bibinfo  {journal} {Phys. Rev.}\ }\textbf {\bibinfo
  {volume} {D82}},\ \bibinfo {pages} {084035} (\bibinfo {year}
  {2010}{\natexlab{a}})},\ \Eprint {http://arxiv.org/abs/1003.3483}
  {arXiv:1003.3483 [gr-qc]} \BibitemShut {NoStop}%
\bibitem [{\citenamefont {Engle}\ \emph
  {et~al.}(2008{\natexlab{b}})\citenamefont {Engle}, \citenamefont {Pereira},\
  and\ \citenamefont {Rovelli}}]{Engle:2007qf}%
  \BibitemOpen
  \bibfield  {author} {\bibinfo {author} {\bibfnamefont {Jonathan}\
  \bibnamefont {Engle}}, \bibinfo {author} {\bibfnamefont {Roberto}\
  \bibnamefont {Pereira}}, \ and\ \bibinfo {author} {\bibfnamefont {Carlo}\
  \bibnamefont {Rovelli}},\ }\bibfield  {title} {\enquote {\bibinfo {title}
  {{Flipped spinfoam vertex and loop gravity}},}\ }\href {\doibase
  10.1016/j.nuclphysb.2008.02.002} {\bibfield  {journal} {\bibinfo  {journal}
  {Nucl.Phys.}\ }\textbf {\bibinfo {volume} {B798}},\ \bibinfo {pages}
  {251--290} (\bibinfo {year} {2008}{\natexlab{b}})},\ \Eprint
  {http://arxiv.org/abs/0708.1236} {arXiv:0708.1236 [gr-qc]} \BibitemShut
  {NoStop}%
\bibitem [{\citenamefont {Conrady}\ and\ \citenamefont
  {Freidel}(2008)}]{Conrady:2008mk}%
  \BibitemOpen
  \bibfield  {author} {\bibinfo {author} {\bibfnamefont {Florian}\ \bibnamefont
  {Conrady}}\ and\ \bibinfo {author} {\bibfnamefont {Laurent}\ \bibnamefont
  {Freidel}},\ }\bibfield  {title} {\enquote {\bibinfo {title} {{On the
  semiclassical limit of 4d spin foam models}},}\ }\href {\doibase
  10.1103/PhysRevD.78.104023} {\bibfield  {journal} {\bibinfo  {journal} {Phys.
  Rev.}\ }\textbf {\bibinfo {volume} {D78}},\ \bibinfo {pages} {104023}
  (\bibinfo {year} {2008})},\ \Eprint {http://arxiv.org/abs/0809.2280}
  {arXiv:0809.2280 [gr-qc]} \BibitemShut {NoStop}%
\bibitem [{\citenamefont {Barrett}\ \emph {et~al.}(2009)\citenamefont
  {Barrett}, \citenamefont {Dowdall}, \citenamefont {Fairbairn}, \citenamefont
  {Gomes},\ and\ \citenamefont {Hellmann}}]{Barrett:2009gg}%
  \BibitemOpen
  \bibfield  {author} {\bibinfo {author} {\bibfnamefont {John~W.}\ \bibnamefont
  {Barrett}}, \bibinfo {author} {\bibfnamefont {Richard~J.}\ \bibnamefont
  {Dowdall}}, \bibinfo {author} {\bibfnamefont {Winston~J.}\ \bibnamefont
  {Fairbairn}}, \bibinfo {author} {\bibfnamefont {Henrique}\ \bibnamefont
  {Gomes}}, \ and\ \bibinfo {author} {\bibfnamefont {Frank}\ \bibnamefont
  {Hellmann}},\ }\bibfield  {title} {\enquote {\bibinfo {title} {{Asymptotic
  analysis of the EPRL four-simplex amplitude}},}\ }\href {\doibase
  10.1063/1.3244218} {\bibfield  {journal} {\bibinfo  {journal} {J. Math.
  Phys.}\ }\textbf {\bibinfo {volume} {50}},\ \bibinfo {pages} {112504}
  (\bibinfo {year} {2009})},\ \Eprint {http://arxiv.org/abs/0902.1170}
  {arXiv:0902.1170 [gr-qc]} \BibitemShut {NoStop}%
\bibitem [{\citenamefont {Barrett}\ \emph {et~al.}(2010)\citenamefont
  {Barrett}, \citenamefont {Dowdall}, \citenamefont {Fairbairn}, \citenamefont
  {Hellmann},\ and\ \citenamefont {Pereira}}]{Barrett:2009mw}%
  \BibitemOpen
  \bibfield  {author} {\bibinfo {author} {\bibfnamefont {John~W.}\ \bibnamefont
  {Barrett}}, \bibinfo {author} {\bibfnamefont {Richard~J.}\ \bibnamefont
  {Dowdall}}, \bibinfo {author} {\bibfnamefont {Winston~J.}\ \bibnamefont
  {Fairbairn}}, \bibinfo {author} {\bibfnamefont {Frank}\ \bibnamefont
  {Hellmann}}, \ and\ \bibinfo {author} {\bibfnamefont {Roberto}\ \bibnamefont
  {Pereira}},\ }\bibfield  {title} {\enquote {\bibinfo {title} {{Lorentzian
  spin foam amplitudes: graphical calculus and asymptotics}},}\ }\href
  {\doibase 10.1088/0264-9381/27/16/165009} {\bibfield  {journal} {\bibinfo
  {journal} {Class. Quant. Grav.}\ }\textbf {\bibinfo {volume} {27}},\ \bibinfo
  {pages} {165009} (\bibinfo {year} {2010})},\ \Eprint
  {http://arxiv.org/abs/0907.2440} {arXiv:0907.2440 [gr-qc]} \BibitemShut
  {NoStop}%
\bibitem [{\citenamefont {Bloch}(1946)}]{Bloch:1946zza}%
  \BibitemOpen
  \bibfield  {author} {\bibinfo {author} {\bibfnamefont {F.}~\bibnamefont
  {Bloch}},\ }\bibfield  {title} {\enquote {\bibinfo {title} {{Nuclear
  Induction}},}\ }\href {\doibase 10.1103/PhysRev.70.460} {\bibfield  {journal}
  {\bibinfo  {journal} {Phys.Rev.}\ }\textbf {\bibinfo {volume} {70}},\
  \bibinfo {pages} {460--474} (\bibinfo {year} {1946})}\BibitemShut {NoStop}%
\bibitem [{\citenamefont {Livine}\ and\ \citenamefont
  {Speziale}(2007)}]{Livine:2007vk}%
  \BibitemOpen
  \bibfield  {author} {\bibinfo {author} {\bibfnamefont {Etera~R.}\
  \bibnamefont {Livine}}\ and\ \bibinfo {author} {\bibfnamefont {Simone}\
  \bibnamefont {Speziale}},\ }\bibfield  {title} {\enquote {\bibinfo {title}
  {{A new spinfoam vertex for quantum gravity}},}\ }\href {\doibase
  10.1103/PhysRevD.76.084028} {\bibfield  {journal} {\bibinfo  {journal} {Phys.
  Rev.}\ }\textbf {\bibinfo {volume} {D76}},\ \bibinfo {pages} {084028}
  (\bibinfo {year} {2007})},\ \Eprint {http://arxiv.org/abs/0705.0674}
  {arXiv:0705.0674 [gr-qc]} \BibitemShut {NoStop}%
\bibitem [{\citenamefont {Bianchi}\ \emph
  {et~al.}(2010{\natexlab{b}})\citenamefont {Bianchi}, \citenamefont {Regoli},\
  and\ \citenamefont {Rovelli}}]{Bianchi:2010fj}%
  \BibitemOpen
  \bibfield  {author} {\bibinfo {author} {\bibfnamefont {Eugenio}\ \bibnamefont
  {Bianchi}}, \bibinfo {author} {\bibfnamefont {Daniele}\ \bibnamefont
  {Regoli}}, \ and\ \bibinfo {author} {\bibfnamefont {Carlo}\ \bibnamefont
  {Rovelli}},\ }\bibfield  {title} {\enquote {\bibinfo {title} {{Face amplitude
  of spinfoam quantum gravity}},}\ }\href {\doibase
  10.1088/0264-9381/27/18/185009} {\bibfield  {journal} {\bibinfo  {journal}
  {Class. Quant. Grav.}\ }\textbf {\bibinfo {volume} {27}},\ \bibinfo {pages}
  {185009} (\bibinfo {year} {2010}{\natexlab{b}})},\ \Eprint
  {http://arxiv.org/abs/1005.0764} {arXiv:1005.0764 [gr-qc]} \BibitemShut
  {NoStop}%
\bibitem [{\citenamefont {Perelomov}()}]{Perelomov:1986tf}%
  \BibitemOpen
  \bibfield  {author} {\bibinfo {author} {\bibfnamefont {A.M.}\ \bibnamefont
  {Perelomov}},\ }\bibfield  {title} {\enquote {\bibinfo {title} {{Generalized
  coherent states and their applications}},}\ }\href@noop {} {\ }\bibinfo
  {note} {Berlin, Germany: Springer (1986) 320 p}\BibitemShut {NoStop}%
\bibitem [{\citenamefont {Markopoulou}(2003)}]{Markopoulou:2002ja}%
  \BibitemOpen
  \bibfield  {author} {\bibinfo {author} {\bibfnamefont {Fotini}\ \bibnamefont
  {Markopoulou}},\ }\bibfield  {title} {\enquote {\bibinfo {title} {{Coarse
  graining in spin foam models}},}\ }\href {\doibase
  10.1088/0264-9381/20/5/301} {\bibfield  {journal} {\bibinfo  {journal}
  {Class.Quant.Grav.}\ }\textbf {\bibinfo {volume} {20}},\ \bibinfo {pages}
  {777--800} (\bibinfo {year} {2003})},\ \Eprint
  {http://arxiv.org/abs/gr-qc/0203036} {arXiv:gr-qc/0203036 [gr-qc]}
  \BibitemShut {NoStop}%
\bibitem [{\citenamefont {Bahr}\ \emph {et~al.}(2011)\citenamefont {Bahr},
  \citenamefont {Dittrich},\ and\ \citenamefont {He}}]{Bahr:2010cq}%
  \BibitemOpen
  \bibfield  {author} {\bibinfo {author} {\bibfnamefont {Benjamin}\
  \bibnamefont {Bahr}}, \bibinfo {author} {\bibfnamefont {Bianca}\ \bibnamefont
  {Dittrich}}, \ and\ \bibinfo {author} {\bibfnamefont {Song}\ \bibnamefont
  {He}},\ }\bibfield  {title} {\enquote {\bibinfo {title} {{Coarse graining
  free theories with gauge symmetries: the linearized case}},}\ }\href
  {\doibase 10.1088/1367-2630/13/4/045009} {\bibfield  {journal} {\bibinfo
  {journal} {New J.Phys.}\ }\textbf {\bibinfo {volume} {13}},\ \bibinfo {pages}
  {045009} (\bibinfo {year} {2011})},\ \Eprint {http://arxiv.org/abs/1011.3667}
  {arXiv:1011.3667 [gr-qc]} \BibitemShut {NoStop}%
\bibitem [{\citenamefont {Daum}\ and\ \citenamefont
  {Reuter}(2012)}]{Daum:2010qt}%
  \BibitemOpen
  \bibfield  {author} {\bibinfo {author} {\bibfnamefont {J.-E.}\ \bibnamefont
  {Daum}}\ and\ \bibinfo {author} {\bibfnamefont {M.}~\bibnamefont {Reuter}},\
  }\bibfield  {title} {\enquote {\bibinfo {title} {{Renormalization Group Flow
  of the Holst Action}},}\ }\href {\doibase 10.1016/j.physletb.2012.01.046}
  {\bibfield  {journal} {\bibinfo  {journal} {Phys.Lett.}\ }\textbf {\bibinfo
  {volume} {B710}},\ \bibinfo {pages} {215--218} (\bibinfo {year} {2012})},\
  \Eprint {http://arxiv.org/abs/1012.4280} {arXiv:1012.4280 [hep-th]}
  \BibitemShut {NoStop}%
\bibitem [{\citenamefont {Benedetti}\ and\ \citenamefont
  {Speziale}(2011)}]{Benedetti:2011nd}%
  \BibitemOpen
  \bibfield  {author} {\bibinfo {author} {\bibfnamefont {Dario}\ \bibnamefont
  {Benedetti}}\ and\ \bibinfo {author} {\bibfnamefont {Simone}\ \bibnamefont
  {Speziale}},\ }\bibfield  {title} {\enquote {\bibinfo {title} {{Perturbative
  quantum gravity with the Immirzi parameter}},}\ }\href {\doibase
  10.1007/JHEP06(2011)107} {\bibfield  {journal} {\bibinfo  {journal} {JHEP}\
  }\textbf {\bibinfo {volume} {1106}},\ \bibinfo {pages} {107} (\bibinfo {year}
  {2011})},\ \Eprint {http://arxiv.org/abs/1104.4028} {arXiv:1104.4028
  [hep-th]} \BibitemShut {NoStop}%
\bibitem [{\citenamefont {Hellmann}()}]{Hellmannprivate}%
  \BibitemOpen
  \bibfield  {author} {\bibinfo {author} {\bibfnamefont {Frank}\ \bibnamefont
  {Hellmann}},\ }\href@noop {} {\ }\bibinfo {note} {Private
  communication}\BibitemShut {NoStop}%
\bibitem [{\citenamefont {Perini}(2012)}]{Perini:2012nd}%
  \BibitemOpen
  \bibfield  {author} {\bibinfo {author} {\bibfnamefont {Claudio}\ \bibnamefont
  {Perini}},\ }\bibfield  {title} {\enquote {\bibinfo {title} {{Holonomy-flux
  spinfoam amplitude}},}\ }\href@noop {} {\  (\bibinfo {year} {2012})},\
  \Eprint {http://arxiv.org/abs/1211.4807} {arXiv:1211.4807 [gr-qc]}
  \BibitemShut {NoStop}%
\bibitem [{\citenamefont {Han}\ and\ \citenamefont
  {Krajewski}(2014)}]{Han:2013gna}%
  \BibitemOpen
  \bibfield  {author} {\bibinfo {author} {\bibfnamefont {Muxin}\ \bibnamefont
  {Han}}\ and\ \bibinfo {author} {\bibfnamefont {Thomas}\ \bibnamefont
  {Krajewski}},\ }\bibfield  {title} {\enquote {\bibinfo {title} {{Path
  Integral Representation of Lorentzian Spinfoam Model, Asymptotics, and
  Simplicial Geometries}},}\ }\href {\doibase 10.1088/0264-9381/31/1/015009}
  {\bibfield  {journal} {\bibinfo  {journal} {Class.Quant.Grav.}\ }\textbf
  {\bibinfo {volume} {31}},\ \bibinfo {pages} {015009} (\bibinfo {year}
  {2014})},\ \Eprint {http://arxiv.org/abs/1304.5626} {arXiv:1304.5626 [gr-qc]}
  \BibitemShut {NoStop}%
\bibitem [{\citenamefont {Barrett}\ and\ \citenamefont
  {Foxon}(1994)}]{Barrett:1993db}%
  \BibitemOpen
  \bibfield  {author} {\bibinfo {author} {\bibfnamefont {John~W.}\ \bibnamefont
  {Barrett}}\ and\ \bibinfo {author} {\bibfnamefont {T.J.}\ \bibnamefont
  {Foxon}},\ }\bibfield  {title} {\enquote {\bibinfo {title} {{Semiclassical
  limits of simplicial quantum gravity}},}\ }\href {\doibase
  10.1088/0264-9381/11/3/009} {\bibfield  {journal} {\bibinfo  {journal}
  {Class.Quant.Grav.}\ }\textbf {\bibinfo {volume} {11}},\ \bibinfo {pages}
  {543--556} (\bibinfo {year} {1994})},\ \Eprint
  {http://arxiv.org/abs/gr-qc/9310016} {arXiv:gr-qc/9310016 [gr-qc]}
  \BibitemShut {NoStop}%
\bibitem [{\citenamefont {Dowdall}\ \emph {et~al.}(2010)\citenamefont
  {Dowdall}, \citenamefont {Gomes},\ and\ \citenamefont
  {Hellmann}}]{Dowdall:2009eg}%
  \BibitemOpen
  \bibfield  {author} {\bibinfo {author} {\bibfnamefont {Richard~J.}\
  \bibnamefont {Dowdall}}, \bibinfo {author} {\bibfnamefont {Henrique}\
  \bibnamefont {Gomes}}, \ and\ \bibinfo {author} {\bibfnamefont {Frank}\
  \bibnamefont {Hellmann}},\ }\bibfield  {title} {\enquote {\bibinfo {title}
  {{Asymptotic analysis of the Ponzano-Regge model for handlebodies}},}\ }\href
  {\doibase 10.1088/1751-8113/43/11/115203} {\bibfield  {journal} {\bibinfo
  {journal} {J.Phys.A}\ }\textbf {\bibinfo {volume} {A43}},\ \bibinfo {pages}
  {115203} (\bibinfo {year} {2010})},\ \Eprint {http://arxiv.org/abs/0909.2027}
  {arXiv:0909.2027 [gr-qc]} \BibitemShut {NoStop}%
\bibitem [{\citenamefont {Bonzom}(2009)}]{Bonzom:2009hw}%
  \BibitemOpen
  \bibfield  {author} {\bibinfo {author} {\bibfnamefont {Valentin}\
  \bibnamefont {Bonzom}},\ }\bibfield  {title} {\enquote {\bibinfo {title}
  {{Spin foam models for quantum gravity from lattice path integrals}},}\
  }\href {\doibase 10.1103/PhysRevD.80.064028} {\bibfield  {journal} {\bibinfo
  {journal} {Phys.Rev.}\ }\textbf {\bibinfo {volume} {D80}},\ \bibinfo {pages}
  {064028} (\bibinfo {year} {2009})},\ \Eprint {http://arxiv.org/abs/0905.1501}
  {arXiv:0905.1501 [gr-qc]} \BibitemShut {NoStop}%
\bibitem [{\citenamefont {Hellmann}\ and\ \citenamefont
  {Kaminski}(2013)}]{Hellmann:2013gva}%
  \BibitemOpen
  \bibfield  {author} {\bibinfo {author} {\bibfnamefont {Frank}\ \bibnamefont
  {Hellmann}}\ and\ \bibinfo {author} {\bibfnamefont {Wojciech}\ \bibnamefont
  {Kaminski}},\ }\bibfield  {title} {\enquote {\bibinfo {title} {{Holonomy spin
  foam models: Asymptotic geometry of the partition function}},}\ }\href
  {\doibase 10.1007/JHEP10(2013)165} {\bibfield  {journal} {\bibinfo  {journal}
  {JHEP}\ }\textbf {\bibinfo {volume} {1310}},\ \bibinfo {pages} {165}
  (\bibinfo {year} {2013})},\ \Eprint {http://arxiv.org/abs/1307.1679}
  {arXiv:1307.1679 [gr-qc]} \BibitemShut {NoStop}%
\bibitem [{\citenamefont {Mamone}\ and\ \citenamefont
  {Rovelli}(2009)}]{Mamone:2009pw}%
  \BibitemOpen
  \bibfield  {author} {\bibinfo {author} {\bibfnamefont {Davide}\ \bibnamefont
  {Mamone}}\ and\ \bibinfo {author} {\bibfnamefont {Carlo}\ \bibnamefont
  {Rovelli}},\ }\bibfield  {title} {\enquote {\bibinfo {title} {{Second-order
  amplitudes in loop quantum gravity}},}\ }\href {\doibase
  10.1088/0264-9381/26/24/245013} {\bibfield  {journal} {\bibinfo  {journal}
  {Class.Quant.Grav.}\ }\textbf {\bibinfo {volume} {26}},\ \bibinfo {pages}
  {245013} (\bibinfo {year} {2009})},\ \Eprint {http://arxiv.org/abs/0904.3730}
  {arXiv:0904.3730 [gr-qc]} \BibitemShut {NoStop}%
\bibitem [{\citenamefont {Han}(2013)}]{Han:2013ina}%
  \BibitemOpen
  \bibfield  {author} {\bibinfo {author} {\bibfnamefont {Muxin}\ \bibnamefont
  {Han}},\ }\bibfield  {title} {\enquote {\bibinfo {title} {{Semiclassical
  Analysis of Spinfoam Model with a Small Barbero-Immirzi Parameter}},}\ }\href
  {\doibase 10.1103/PhysRevD.88.044051} {\bibfield  {journal} {\bibinfo
  {journal} {Phys.Rev.}\ }\textbf {\bibinfo {volume} {D88}},\ \bibinfo {pages}
  {044051} (\bibinfo {year} {2013})},\ \Eprint {http://arxiv.org/abs/1304.5628}
  {arXiv:1304.5628 [gr-qc]} \BibitemShut {NoStop}%
\bibitem [{\citenamefont {Regge}(1961)}]{Regge:1961px}%
  \BibitemOpen
  \bibfield  {author} {\bibinfo {author} {\bibfnamefont {T.}~\bibnamefont
  {Regge}},\ }\bibfield  {title} {\enquote {\bibinfo {title} {{General
  relativity without coordinates}},}\ }\href {\doibase 10.1007/BF02733251}
  {\bibfield  {journal} {\bibinfo  {journal} {Nuovo Cim.}\ }\textbf {\bibinfo
  {volume} {19}},\ \bibinfo {pages} {558--571} (\bibinfo {year}
  {1961})}\BibitemShut {NoStop}%
\bibitem [{\citenamefont {Williams}\ and\ \citenamefont
  {Tuckey}(1992)}]{Williams:1991cd}%
  \BibitemOpen
  \bibfield  {author} {\bibinfo {author} {\bibfnamefont {Ruth~M.}\ \bibnamefont
  {Williams}}\ and\ \bibinfo {author} {\bibfnamefont {Philip~A.}\ \bibnamefont
  {Tuckey}},\ }\bibfield  {title} {\enquote {\bibinfo {title} {{Regge calculus:
  A Bibliography and brief review}},}\ }\href {\doibase
  10.1088/0264-9381/9/5/021} {\bibfield  {journal} {\bibinfo  {journal}
  {Class.Quant.Grav.}\ }\textbf {\bibinfo {volume} {9}},\ \bibinfo {pages}
  {1409--1422} (\bibinfo {year} {1992})}\BibitemShut {NoStop}%
\bibitem [{\citenamefont {Barrett}\ \emph {et~al.}(1999)\citenamefont
  {Barrett}, \citenamefont {Rocek},\ and\ \citenamefont
  {Williams}}]{Barrett:1997tx}%
  \BibitemOpen
  \bibfield  {author} {\bibinfo {author} {\bibfnamefont {John~W.}\ \bibnamefont
  {Barrett}}, \bibinfo {author} {\bibfnamefont {Martin}\ \bibnamefont {Rocek}},
  \ and\ \bibinfo {author} {\bibfnamefont {Ruth~M.}\ \bibnamefont {Williams}},\
  }\bibfield  {title} {\enquote {\bibinfo {title} {{A Note on area variables in
  Regge calculus}},}\ }\href {\doibase 10.1088/0264-9381/16/4/025} {\bibfield
  {journal} {\bibinfo  {journal} {Class.Quant.Grav.}\ }\textbf {\bibinfo
  {volume} {16}},\ \bibinfo {pages} {1373--1376} (\bibinfo {year} {1999})},\
  \Eprint {http://arxiv.org/abs/gr-qc/9710056} {arXiv:gr-qc/9710056 [gr-qc]}
  \BibitemShut {NoStop}%
\bibitem [{\citenamefont {Makela}\ and\ \citenamefont
  {Williams}(2001)}]{Makela:2000ej}%
  \BibitemOpen
  \bibfield  {author} {\bibinfo {author} {\bibfnamefont {Jarmo}\ \bibnamefont
  {Makela}}\ and\ \bibinfo {author} {\bibfnamefont {Ruth~M.}\ \bibnamefont
  {Williams}},\ }\bibfield  {title} {\enquote {\bibinfo {title} {{Constraints
  on area variables in Regge calculus}},}\ }\href {\doibase
  10.1088/0264-9381/18/4/102} {\bibfield  {journal} {\bibinfo  {journal}
  {Class.Quant.Grav.}\ }\textbf {\bibinfo {volume} {18}},\ \bibinfo {pages}
  {L43} (\bibinfo {year} {2001})},\ \Eprint
  {http://arxiv.org/abs/gr-qc/0011006} {arXiv:gr-qc/0011006 [gr-qc]}
  \BibitemShut {NoStop}%
\bibitem [{\citenamefont {Dittrich}\ and\ \citenamefont
  {Speziale}(2008)}]{Dittrich:2008va}%
  \BibitemOpen
  \bibfield  {author} {\bibinfo {author} {\bibfnamefont {Bianca}\ \bibnamefont
  {Dittrich}}\ and\ \bibinfo {author} {\bibfnamefont {Simone}\ \bibnamefont
  {Speziale}},\ }\bibfield  {title} {\enquote {\bibinfo {title} {{Area-angle
  variables for general relativity}},}\ }\href {\doibase
  10.1088/1367-2630/10/8/083006} {\bibfield  {journal} {\bibinfo  {journal}
  {New J.Phys.}\ }\textbf {\bibinfo {volume} {10}},\ \bibinfo {pages} {083006}
  (\bibinfo {year} {2008})},\ \Eprint {http://arxiv.org/abs/0802.0864}
  {arXiv:0802.0864 [gr-qc]} \BibitemShut {NoStop}%
\bibitem [{\citenamefont {Rocek}\ and\ \citenamefont
  {Williams}(1981)}]{Rocek:1982fr}%
  \BibitemOpen
  \bibfield  {author} {\bibinfo {author} {\bibfnamefont {M.}~\bibnamefont
  {Rocek}}\ and\ \bibinfo {author} {\bibfnamefont {Ruth~M.}\ \bibnamefont
  {Williams}},\ }\bibfield  {title} {\enquote {\bibinfo {title} {{QUANTUM REGGE
  CALCULUS}},}\ }\href {\doibase 10.1016/0370-2693(81)90848-0} {\bibfield
  {journal} {\bibinfo  {journal} {Phys.Lett.}\ }\textbf {\bibinfo {volume}
  {B104}},\ \bibinfo {pages} {31} (\bibinfo {year} {1981})}\BibitemShut
  {NoStop}%
\bibitem [{\citenamefont {Engle}\ and\ \citenamefont
  {Pereira}(2009)}]{Engle:2008ev}%
  \BibitemOpen
  \bibfield  {author} {\bibinfo {author} {\bibfnamefont {Jonathan}\
  \bibnamefont {Engle}}\ and\ \bibinfo {author} {\bibfnamefont {Roberto}\
  \bibnamefont {Pereira}},\ }\bibfield  {title} {\enquote {\bibinfo {title}
  {{Regularization and finiteness of the Lorentzian LQG vertices}},}\ }\href
  {\doibase 10.1103/PhysRevD.79.084034} {\bibfield  {journal} {\bibinfo
  {journal} {Phys.Rev.}\ }\textbf {\bibinfo {volume} {D79}},\ \bibinfo {pages}
  {084034} (\bibinfo {year} {2009})},\ \Eprint {http://arxiv.org/abs/0805.4696}
  {arXiv:0805.4696 [gr-qc]} \BibitemShut {NoStop}%
\bibitem [{\citenamefont {Ashtekar}\ \emph {et~al.}(2010)\citenamefont
  {Ashtekar}, \citenamefont {Campiglia},\ and\ \citenamefont
  {Henderson}}]{Ashtekar:2010gz}%
  \BibitemOpen
  \bibfield  {author} {\bibinfo {author} {\bibfnamefont {Abhay}\ \bibnamefont
  {Ashtekar}}, \bibinfo {author} {\bibfnamefont {Miguel}\ \bibnamefont
  {Campiglia}}, \ and\ \bibinfo {author} {\bibfnamefont {Adam}\ \bibnamefont
  {Henderson}},\ }\bibfield  {title} {\enquote {\bibinfo {title} {{Path
  Integrals and the WKB approximation in Loop Quantum Cosmology}},}\ }\href
  {\doibase 10.1103/PhysRevD.82.124043} {\bibfield  {journal} {\bibinfo
  {journal} {Phys.Rev.}\ }\textbf {\bibinfo {volume} {D82}},\ \bibinfo {pages}
  {124043} (\bibinfo {year} {2010})},\ \Eprint {http://arxiv.org/abs/1011.1024}
  {arXiv:1011.1024 [gr-qc]} \BibitemShut {NoStop}%
\bibitem [{\citenamefont {Noui}\ and\ \citenamefont
  {Perez}(2005)}]{Noui:2004iy}%
  \BibitemOpen
  \bibfield  {author} {\bibinfo {author} {\bibfnamefont {Karim}\ \bibnamefont
  {Noui}}\ and\ \bibinfo {author} {\bibfnamefont {Alejandro}\ \bibnamefont
  {Perez}},\ }\bibfield  {title} {\enquote {\bibinfo {title} {{Three
  dimensional loop quantum gravity: Physical scalar product and spin foam
  models}},}\ }\href {\doibase 10.1088/0264-9381/22/9/017} {\bibfield
  {journal} {\bibinfo  {journal} {Class. Quant. Grav.}\ }\textbf {\bibinfo
  {volume} {22}},\ \bibinfo {pages} {1739--1762} (\bibinfo {year} {2005})},\
  \Eprint {http://arxiv.org/abs/gr-qc/0402110} {arXiv:gr-qc/0402110}
  \BibitemShut {NoStop}%
\bibitem [{\citenamefont {Oeckl}(2003)}]{Oeckl:2003vu}%
  \BibitemOpen
  \bibfield  {author} {\bibinfo {author} {\bibfnamefont {Robert}\ \bibnamefont
  {Oeckl}},\ }\bibfield  {title} {\enquote {\bibinfo {title} {{A 'general
  boundary' formulation for quantum mechanics and quantum gravity}},}\ }\href
  {\doibase 10.1016/j.physletb.2003.08.043} {\bibfield  {journal} {\bibinfo
  {journal} {Phys. Lett.}\ }\textbf {\bibinfo {volume} {B575}},\ \bibinfo
  {pages} {318--324} (\bibinfo {year} {2003})},\ \Eprint
  {http://arxiv.org/abs/hep-th/0306025} {arXiv:hep-th/0306025} \BibitemShut
  {NoStop}%
\bibitem [{\citenamefont {Bianchi}\ and\ \citenamefont
  {Satz}(2009)}]{Bianchi:2008ae}%
  \BibitemOpen
  \bibfield  {author} {\bibinfo {author} {\bibfnamefont {Eugenio}\ \bibnamefont
  {Bianchi}}\ and\ \bibinfo {author} {\bibfnamefont {Alejandro}\ \bibnamefont
  {Satz}},\ }\bibfield  {title} {\enquote {\bibinfo {title} {{Semiclassical
  regime of Regge calculus and spin foams}},}\ }\href {\doibase
  10.1016/j.nuclphysb.2008.09.005} {\bibfield  {journal} {\bibinfo  {journal}
  {Nucl.Phys.}\ }\textbf {\bibinfo {volume} {B808}},\ \bibinfo {pages}
  {546--568} (\bibinfo {year} {2009})},\ \Eprint
  {http://arxiv.org/abs/0808.1107} {arXiv:0808.1107 [gr-qc]} \BibitemShut
  {NoStop}%
\bibitem [{\citenamefont {Sorkin}(1975)}]{Sorkin:1975ah}%
  \BibitemOpen
  \bibfield  {author} {\bibinfo {author} {\bibfnamefont {R.}~\bibnamefont
  {Sorkin}},\ }\bibfield  {title} {\enquote {\bibinfo {title} {{Time Evolution
  Problem in Regge Calculus}},}\ }\href {\doibase 10.1103/PhysRevD.23.565,
  10.1103/PhysRevD.12.385} {\bibfield  {journal} {\bibinfo  {journal}
  {Phys.Rev.}\ }\textbf {\bibinfo {volume} {D12}},\ \bibinfo {pages} {385--396}
  (\bibinfo {year} {1975})}\BibitemShut {NoStop}%
\bibitem [{\citenamefont {Barrett}\ \emph {et~al.}(1997)\citenamefont
  {Barrett}, \citenamefont {Galassi}, \citenamefont {Miller}, \citenamefont
  {Sorkin}, \citenamefont {Tuckey} \emph {et~al.}}]{Barrett:1994ks}%
  \BibitemOpen
  \bibfield  {author} {\bibinfo {author} {\bibfnamefont {John~W.}\ \bibnamefont
  {Barrett}}, \bibinfo {author} {\bibfnamefont {Mark}\ \bibnamefont {Galassi}},
  \bibinfo {author} {\bibfnamefont {Warner~A.}\ \bibnamefont {Miller}},
  \bibinfo {author} {\bibfnamefont {Rafael~D.}\ \bibnamefont {Sorkin}},
  \bibinfo {author} {\bibfnamefont {Philip~A.}\ \bibnamefont {Tuckey}},  \emph
  {et~al.},\ }\bibfield  {title} {\enquote {\bibinfo {title} {{A Paralellizable
  implicit evolution scheme for Regge calculus}},}\ }\href {\doibase
  10.1007/BF02435787} {\bibfield  {journal} {\bibinfo  {journal}
  {Int.J.Theor.Phys.}\ }\textbf {\bibinfo {volume} {36}},\ \bibinfo {pages}
  {815--840} (\bibinfo {year} {1997})},\ \Eprint
  {http://arxiv.org/abs/gr-qc/9411008} {arXiv:gr-qc/9411008 [gr-qc]}
  \BibitemShut {NoStop}%
\bibitem [{\citenamefont {Bianchi}\ and\ \citenamefont
  {Ding}(2012)}]{Bianchi:2011hp}%
  \BibitemOpen
  \bibfield  {author} {\bibinfo {author} {\bibfnamefont {Eugenio}\ \bibnamefont
  {Bianchi}}\ and\ \bibinfo {author} {\bibfnamefont {You}\ \bibnamefont
  {Ding}},\ }\bibfield  {title} {\enquote {\bibinfo {title} {{Lorentzian
  spinfoam propagator}},}\ }\href {\doibase 10.1103/PhysRevD.86.104040}
  {\bibfield  {journal} {\bibinfo  {journal} {Phys.Rev.}\ }\textbf {\bibinfo
  {volume} {D86}},\ \bibinfo {pages} {104040} (\bibinfo {year} {2012})},\
  \Eprint {http://arxiv.org/abs/1109.6538} {arXiv:1109.6538 [gr-qc]}
  \BibitemShut {NoStop}%
\bibitem [{\citenamefont {Rovelli}\ and\ \citenamefont
  {Zhang}(2011)}]{Rovelli:2011kf}%
  \BibitemOpen
  \bibfield  {author} {\bibinfo {author} {\bibfnamefont {Carlo}\ \bibnamefont
  {Rovelli}}\ and\ \bibinfo {author} {\bibfnamefont {Mingyi}\ \bibnamefont
  {Zhang}},\ }\bibfield  {title} {\enquote {\bibinfo {title} {{Euclidean
  three-point function in loop and perturbative gravity}},}\ }\href {\doibase
  10.1088/0264-9381/28/17/175010} {\bibfield  {journal} {\bibinfo  {journal}
  {Class.Quant.Grav.}\ }\textbf {\bibinfo {volume} {28}},\ \bibinfo {pages}
  {175010} (\bibinfo {year} {2011})},\ \Eprint {http://arxiv.org/abs/1105.0566}
  {arXiv:1105.0566 [gr-qc]} \BibitemShut {NoStop}%
\bibitem [{\citenamefont {Bojowald}(2001)}]{Bojowald:2001ep}%
  \BibitemOpen
  \bibfield  {author} {\bibinfo {author} {\bibfnamefont {Martin}\ \bibnamefont
  {Bojowald}},\ }\bibfield  {title} {\enquote {\bibinfo {title} {{The
  semiclassical limit of loop quantum cosmology}},}\ }\href {\doibase
  10.1088/0264-9381/18/18/101} {\bibfield  {journal} {\bibinfo  {journal}
  {Class. Quant. Grav.}\ }\textbf {\bibinfo {volume} {18}},\ \bibinfo {pages}
  {L109--L116} (\bibinfo {year} {2001})},\ \Eprint
  {http://arxiv.org/abs/gr-qc/0105113} {arXiv:gr-qc/0105113} \BibitemShut
  {NoStop}%
\bibitem [{\citenamefont {Bianchi}\ \emph
  {et~al.}(2010{\natexlab{c}})\citenamefont {Bianchi}, \citenamefont
  {Magliaro},\ and\ \citenamefont {Perini}}]{Bianchi:2010mw}%
  \BibitemOpen
  \bibfield  {author} {\bibinfo {author} {\bibfnamefont {Eugenio}\ \bibnamefont
  {Bianchi}}, \bibinfo {author} {\bibfnamefont {Elena}\ \bibnamefont
  {Magliaro}}, \ and\ \bibinfo {author} {\bibfnamefont {Claudio}\ \bibnamefont
  {Perini}},\ }\bibfield  {title} {\enquote {\bibinfo {title} {{Spinfoams in
  the holomorphic representation}},}\ }\href {\doibase
  10.1103/PhysRevD.82.124031} {\bibfield  {journal} {\bibinfo  {journal}
  {Phys.Rev.}\ }\textbf {\bibinfo {volume} {D82}},\ \bibinfo {pages} {124031}
  (\bibinfo {year} {2010}{\natexlab{c}})},\ \Eprint
  {http://arxiv.org/abs/1004.4550} {arXiv:1004.4550 [gr-qc]} \BibitemShut
  {NoStop}%
\bibitem [{\citenamefont {Livine}\ and\ \citenamefont
  {Oriti}(2003)}]{Livine:2002rh}%
  \BibitemOpen
  \bibfield  {author} {\bibinfo {author} {\bibfnamefont {Etera~R.}\
  \bibnamefont {Livine}}\ and\ \bibinfo {author} {\bibfnamefont {Daniele}\
  \bibnamefont {Oriti}},\ }\bibfield  {title} {\enquote {\bibinfo {title}
  {{Implementing causality in the spin foam quantum geometry}},}\ }\href
  {\doibase 10.1016/S0550-3213(03)00378-X} {\bibfield  {journal} {\bibinfo
  {journal} {Nucl.Phys.}\ }\textbf {\bibinfo {volume} {B663}},\ \bibinfo
  {pages} {231--279} (\bibinfo {year} {2003})},\ \Eprint
  {http://arxiv.org/abs/gr-qc/0210064} {arXiv:gr-qc/0210064 [gr-qc]}
  \BibitemShut {NoStop}%
\bibitem [{\citenamefont {Oriti}(2005)}]{Oriti:2004yu}%
  \BibitemOpen
  \bibfield  {author} {\bibinfo {author} {\bibfnamefont {Daniele}\ \bibnamefont
  {Oriti}},\ }\bibfield  {title} {\enquote {\bibinfo {title} {{The Feynman
  propagator for quantum gravity: Spin foams, proper time, orientation,
  causality and timeless-ordering}},}\ }\href@noop {} {\bibfield  {journal}
  {\bibinfo  {journal} {Braz.J.Phys.}\ }\textbf {\bibinfo {volume} {35}},\
  \bibinfo {pages} {481--488} (\bibinfo {year} {2005})},\ \Eprint
  {http://arxiv.org/abs/gr-qc/0412035} {arXiv:gr-qc/0412035 [gr-qc]}
  \BibitemShut {NoStop}%
\bibitem [{\citenamefont {Oriti}\ and\ \citenamefont
  {Tlas}(2006)}]{Oriti:2006wq}%
  \BibitemOpen
  \bibfield  {author} {\bibinfo {author} {\bibfnamefont {Daniele}\ \bibnamefont
  {Oriti}}\ and\ \bibinfo {author} {\bibfnamefont {Tamer}\ \bibnamefont
  {Tlas}},\ }\bibfield  {title} {\enquote {\bibinfo {title} {{Causality and
  matter propagation in 3-D spin foam quantum gravity}},}\ }\href {\doibase
  10.1103/PhysRevD.74.104021} {\bibfield  {journal} {\bibinfo  {journal}
  {Phys.Rev.}\ }\textbf {\bibinfo {volume} {D74}},\ \bibinfo {pages} {104021}
  (\bibinfo {year} {2006})},\ \Eprint {http://arxiv.org/abs/gr-qc/0608116}
  {arXiv:gr-qc/0608116 [gr-qc]} \BibitemShut {NoStop}%
\bibitem [{\citenamefont {Perini}\ \emph {et~al.}(2009)\citenamefont {Perini},
  \citenamefont {Rovelli},\ and\ \citenamefont {Speziale}}]{Perini:2008pd}%
  \BibitemOpen
  \bibfield  {author} {\bibinfo {author} {\bibfnamefont {Claudio}\ \bibnamefont
  {Perini}}, \bibinfo {author} {\bibfnamefont {Carlo}\ \bibnamefont {Rovelli}},
  \ and\ \bibinfo {author} {\bibfnamefont {Simone}\ \bibnamefont {Speziale}},\
  }\bibfield  {title} {\enquote {\bibinfo {title} {{Self-energy and vertex
  radiative corrections in LQG}},}\ }\href {\doibase
  10.1016/j.physletb.2009.10.076} {\bibfield  {journal} {\bibinfo  {journal}
  {Phys. Lett.}\ }\textbf {\bibinfo {volume} {B682}},\ \bibinfo {pages}
  {78--84} (\bibinfo {year} {2009})},\ \Eprint {http://arxiv.org/abs/0810.1714}
  {arXiv:0810.1714 [gr-qc]} \BibitemShut {NoStop}%
\bibitem [{\citenamefont {Krajewski}\ \emph {et~al.}(2010)\citenamefont
  {Krajewski}, \citenamefont {Magnen}, \citenamefont {Rivasseau}, \citenamefont
  {Tanasa},\ and\ \citenamefont {Vitale}}]{Krajewski:2010yq}%
  \BibitemOpen
  \bibfield  {author} {\bibinfo {author} {\bibfnamefont {Thomas}\ \bibnamefont
  {Krajewski}}, \bibinfo {author} {\bibfnamefont {Jacques}\ \bibnamefont
  {Magnen}}, \bibinfo {author} {\bibfnamefont {Vincent}\ \bibnamefont
  {Rivasseau}}, \bibinfo {author} {\bibfnamefont {Adrian}\ \bibnamefont
  {Tanasa}}, \ and\ \bibinfo {author} {\bibfnamefont {Patrizia}\ \bibnamefont
  {Vitale}},\ }\bibfield  {title} {\enquote {\bibinfo {title} {{Quantum
  Corrections in the Group Field Theory Formulation of the EPRL/FK Models}},}\
  }\href {\doibase 10.1103/PhysRevD.82.124069} {\bibfield  {journal} {\bibinfo
  {journal} {Phys.Rev.}\ }\textbf {\bibinfo {volume} {D82}},\ \bibinfo {pages}
  {124069} (\bibinfo {year} {2010})},\ \Eprint {http://arxiv.org/abs/1007.3150}
  {arXiv:1007.3150 [gr-qc]} \BibitemShut {NoStop}%
\bibitem [{\citenamefont {Ben~Geloun}\ \emph {et~al.}(2010)\citenamefont
  {Ben~Geloun}, \citenamefont {Gurau},\ and\ \citenamefont
  {Rivasseau}}]{Geloun:2010vj}%
  \BibitemOpen
  \bibfield  {author} {\bibinfo {author} {\bibfnamefont {Joseph}\ \bibnamefont
  {Ben~Geloun}}, \bibinfo {author} {\bibfnamefont {Razvan}\ \bibnamefont
  {Gurau}}, \ and\ \bibinfo {author} {\bibfnamefont {Vincent}\ \bibnamefont
  {Rivasseau}},\ }\bibfield  {title} {\enquote {\bibinfo {title} {{EPRL/FK
  Group Field Theory}},}\ }\href {\doibase 10.1209/0295-5075/92/60008}
  {\bibfield  {journal} {\bibinfo  {journal} {Europhys.Lett.}\ }\textbf
  {\bibinfo {volume} {92}},\ \bibinfo {pages} {60008} (\bibinfo {year}
  {2010})},\ \Eprint {http://arxiv.org/abs/1008.0354} {arXiv:1008.0354
  [hep-th]} \BibitemShut {NoStop}%
\bibitem [{\citenamefont {Rivasseau}(2010)}]{Rivasseau:2011xg}%
  \BibitemOpen
  \bibfield  {author} {\bibinfo {author} {\bibfnamefont {Vincent}\ \bibnamefont
  {Rivasseau}},\ }\bibfield  {title} {\enquote {\bibinfo {title} {{Towards
  Renormalizing Group Field Theory}},}\ }\href@noop {} {\bibfield  {journal}
  {\bibinfo  {journal} {PoS}\ }\textbf {\bibinfo {volume} {CNCFG2010}},\
  \bibinfo {pages} {004} (\bibinfo {year} {2010})},\ \Eprint
  {http://arxiv.org/abs/1103.1900} {arXiv:1103.1900 [gr-qc]} \BibitemShut
  {NoStop}%
\bibitem [{\citenamefont {Han}(2011)}]{Han:2010pz}%
  \BibitemOpen
  \bibfield  {author} {\bibinfo {author} {\bibfnamefont {Muxin}\ \bibnamefont
  {Han}},\ }\bibfield  {title} {\enquote {\bibinfo {title} {{4-dimensional
  Spin-foam Model with Quantum Lorentz Group}},}\ }\href {\doibase
  10.1063/1.3606592} {\bibfield  {journal} {\bibinfo  {journal} {J.Math.Phys.}\
  }\textbf {\bibinfo {volume} {52}},\ \bibinfo {pages} {072501} (\bibinfo
  {year} {2011})},\ \Eprint {http://arxiv.org/abs/1012.4216} {arXiv:1012.4216
  [gr-qc]} \BibitemShut {NoStop}%
\bibitem [{\citenamefont {Fairbairn}\ and\ \citenamefont
  {Meusburger}(2011)}]{Fairbairn:2011aa}%
  \BibitemOpen
  \bibfield  {author} {\bibinfo {author} {\bibfnamefont {Winston~J.}\
  \bibnamefont {Fairbairn}}\ and\ \bibinfo {author} {\bibfnamefont {Catherine}\
  \bibnamefont {Meusburger}},\ }\bibfield  {title} {\enquote {\bibinfo {title}
  {{q-Deformation of Lorentzian spin foam models}},}\ }\href@noop {} {\bibfield
   {journal} {\bibinfo  {journal} {PoS}\ }\textbf {\bibinfo {volume}
  {QGQGS2011}},\ \bibinfo {pages} {017} (\bibinfo {year} {2011})},\ \Eprint
  {http://arxiv.org/abs/1112.2511} {arXiv:1112.2511 [gr-qc]} \BibitemShut
  {NoStop}%
\end{thebibliography}%
\end{document}